\documentclass[a4paper,fleqn, usenatbib]{mnras}

\usepackage{newtxtext,newtxmath}

\usepackage[T1]{fontenc}
\usepackage{ae,aecompl}

\usepackage{stackengine, graphicx}	
\usepackage{amsmath}	
\usepackage{amssymb}	
\usepackage{lscape}
\usepackage{geometry}
\usepackage{comment}
\usepackage{gensymb}
\usepackage{float}
\usepackage{tablefootnote}

\newcommand{\Cl}{Cl~J1449+0856}
\newcommand{\Htwo}{H$_{2}$}
\newcommand{\kms}{kms$^{-1}$}
\newcommand{\Halpha}{H$\alpha$}
\newcommand{\aco}{$\alpha_{\rm CO}$}
\def\Put(#1,#2)#3{\leavevmode\makebox(0,0){\put(#1,#2){#3}}}

\title[Effects of low metallicity at z=1.99]{Suppressed CO emission and high G/D ratios in z=2 galaxies with sub-solar gas-phase metallicity}

\author[R. T. Coogan et al.]{R. T. Coogan$^{1,2}$$\thanks{E-mail: r.coogan@sussex.ac.uk}$, M. T. Sargent$^{1}$, E. Daddi$^{2}$, F. Valentino$^{3}$, V. Strazzullo$^{4}$, \newauthor M. B\'ethermin$^{5, 6}$,
R. Gobat$^{7}$, D. Liu$^{8}$, G. Magdis$^{3,9}$
\\
$^{1}$Astronomy Centre, Department of Physics and Astronomy, University of Sussex, Brighton BN1 9QH, UK\\
$^{2}$CEA, IRFU, DAp, AIM, Universit\'{e} Paris-Saclay, Universit\'{e} Paris Diderot, Sorbonne Paris Cit\'{e}, CNRS, F-91191 Gif-sur-Yvette, France\\
$^{3}$Cosmic Dawn Center, Niels Bohr Institute, University of Copenhagen, Juliane Mariesvej 30, 2100, Copenhagen, Denmark\\
$^{4}$Department of Physics, Ludwig-Maximilians-Universitat, Scheinerstr.
1, 81679 Munchen, Germany\\
$^{5}$Aix Marseille Univ, CNRS, LAM, Laboratoire d'Astrophysique de Marseille, Marseille, France\\
$^{6}$European Southern Observatory, Karl-Schwarzschild-Str. 2, 85748 Garching, Germany\\
$^{7}$Instituto de F\'{i}sica, Pontificia Universidad Cat\'{o}lica de Valpara\'{i}so, Casilla 4059, Valpara\'{i}so, Chile\\
$^{8}$Max Planck Institute for Astronomy, Konigstuhl 17, D-69117 Heidelberg,
Germany\\
$^{9}$Institute for Astronomy, Astrophysics, Space Applications and Remote Sensing, National Observatory of Athens, 15236, Athens, Greece\\
}

\date{Accepted 2019 January 17. Received 2018 December 22; in original form 2018 September 1.}

\pubyear{2018}

\begin{document}
\label{firstpage}
\pagerange{\pageref{firstpage}--\pageref{lastpage}}
\maketitle

\begin{abstract}
We study a population of significantly sub-solar enrichment galaxies at z=1.99, to investigate how molecular gas, dust and star-formation relate in low-metallicity galaxies at the peak epoch of star-formation. We target our sample with several deep ALMA and VLA datasets, and find no individual detections of CO[4-3], CO[1-0] or dust, in stark contrast to the $>$60\% detection rate expected for solar-enrichment galaxies with these MS \Halpha\ SFRs. We find that both low and high density molecular gas (traced by CO[1-0] and CO[4-3] respectively) are affected by the low enrichment, showing sample average (stacked) luminosity deficits $>$0.5-0.7~dex below expectations. This is particularly pertinent for the use of high-J CO emission as a proxy of instantaneous star-formation rate. Our individual galaxy data and stacked constraints point to a strong inverse dependence $\propto Z^{\gamma}$ of gas-to-dust ratios (G/D) and CO-to-H$_{2}$ conversion factors (\aco) on metallicity at z$\sim$2, with $\gamma_{\rm G/D}<$-2.2 and $\gamma_{\alpha_{\rm CO}}<$-0.8, respectively. We quantify the importance of comparing G/D and \aco\ vs. metallicity trends from the literature on a common, suitably normalised metallicity scale. When accounting for systematic offsets between different metallicity scales, our z$\sim$2 constraints on these scaling relations are consistent with the corresponding relations for local galaxies. However, among those local relations, we favour those with a steep/double power-law dependence of G/D on metallicity. Finally, we discuss the implications of these findings for (a) gas mass measurements for sub-M$^{*}$ galaxies, and (b) efforts to identify the characteristic galaxy mass scale contributing most to the comoving molecular gas density at z=2.

\end{abstract}

\begin{keywords}
galaxies: high-redshift -- galaxies: evolution -- galaxies: ISM -- galaxies: star formation -- galaxies: abundances
\end{keywords}



\section{Introduction}

Our current understanding of galaxy properties at the peak epoch of star-formation (z$\sim$2), in particular the interstellar medium (ISM), is primarily driven by observations of the brightest, most massive galaxies. The interplay between gas, dust and star-formation for the `normal' galaxy population with M$_{\star}\sim$10$^{10}$-10$^{11}$M$_{\odot}$ and approximately solar enrichment at z=2 is now reasonably well characterised (e.g. \citealt{ref:M.Aravena2010, ref:E.Daddi2010a, ref:E.Daddi2010b, ref:R.Genzel2010, ref:G.Magdis2012, ref:L.J.Tacconi2013, ref:L.Tacconi2018, ref:P.Santini2014, ref:M.Sargent2014, ref:N.Scoville2014, ref:N.Scoville2017, ref:Q.Tan2014, ref:R.Decarli2016}). However, the ISM properties of lower mass galaxies at this redshift are much less well-understood, despite the fact that they are far more numerous than higher mass systems. This gap in our knowledge is a consequence of sub-M$^{*}$ galaxies being much more difficult to observe, in part due to their relative faintness. It is however important to push down to these systems to improve our understanding of the gas cycle and consumption across a broad range of halo and stellar masses.

A compounding difficulty in ISM observations is due to the low gas-phase metal abundance in these systems, as described by the `mass-metallicity relation' (MZR), both in the local Universe and at higher redshifts (e.g. \citealt{ref:C.Tremonti2004},  \citealt{ref:H.Zahid2014a}). The low enrichment of these galaxies affects quantities commonly used to estimate ISM content from observational data, such as the gas-to-dust ratio (G/D) and CO-to-\Htwo\ conversion factor (\aco). Molecular gas masses can be estimated from observations through CO spectroscopy, using \aco\ to convert between CO luminosity and molecular gas mass. However, at low oxygen abundance, increasing proportions of star-formation in a galaxy are predicted to occur in `CO-dark' molecular Hydrogen gas (e.g. \citealt{ref:F.Israel1997, ref:I.Grenier2005, ref:M.Wolfire2010}). At low metallicity, the shielding of CO molecules by dust molecules is reduced, leading to higher photo-dissociation of CO molecules by ultra-violet (UV) radiation and a lower CO emission per unit molecular gas mass, thus increasing the value of \aco. 

In addition to this intrinsic lack of CO, the low-J CO transitions become observationally too faint with increasing redshift, and many low-J transitions fall out of suitable observing windows. It is therefore more practical to use sensitive instruments such as the Atacama Large Millimeter/submillimeter Array (ALMA) to detect emission from the higher order transitions of CO. An excitation correction is then applied, thereby adding further assumptions to the task of measuring molecular gas masses. Alternatively, gas masses can be inferred through cold dust continuum observations. This is done by relating the dust emission to the total gas mass through the G/D ratio (e.g. \citealt{ref:M.Guelin1993, ref:E.Corbelli2012, ref:S.Eales2012, ref:G.Magdis2012, ref:N.Bourne2013}, \citealt{ref:N.Scoville2014, ref:B.Groves2015}), which is again closely linked to the metallicity of a galaxy.

Previous studies have therefore largely focussed on characterising the relationship between gas, dust and metallicity of the ISM in local galaxies (e.g. \citealt{ref:D.Burnstein1982}, \citealt{ref:M.Issa1990},  \citealt{ref:B.Draine2007}, \mbox{\citealt{ref:A.Leroy2011a}}, \citealt{ref:G.Magdis2012}, \mbox{\citealt{ref:K.Sandstrom2013}}, \citealt{ref:Q.Tan2014, ref:A.RemyRuyer2014, ref:F.Belfiore2015}, \citealt{ref:R.Amorin2016}). These works have illustrated a clear inverse dependence of G/D and \aco\ on metallicity. However, observations of the high redshift, low-metallicity regime are now required, in order to investigate whether similar relations as observed at low redshift also hold for galaxies in the peak epoch of galaxy formation. With the lack of equivalent information at high redshift, we are currently forced to either assume no metallicity dependence when deriving gas content from continuum or line observations at high redshift, or to assume that the scaling relations calibrated locally are valid at higher redshift. Some early attempts have been made to constrain these scaling relations at high redshift, such as those by \citet{ref:R.Genzel2012} and \citet{ref:A.Saintonge2013}. \citet{ref:R.Genzel2012} used the CO[3-2] and CO[2-1] transitions to perform a study of normal star-forming galaxies between z$\sim$1-2, with stellar mass M$_{\star}\gtrsim$10$^{10}$M$_{\odot}$ and oxygen abundance on the \citet[][D02]{ref:D02} scale of 8.4-8.9. They found a strong inverse dependence of \aco\ on metallicity in star-forming galaxies at z$>$1, similar to what is observed at z$\sim$0. \citet{ref:A.Saintonge2013} study the relationship between G/D and metallicity using $\sim$10$^{10}$M$_{\odot}$ z$>$2 lensed galaxies over a similar metallicity range, and find an increase in the normalisation of the G/D ratio a factor 1.7 above the relation found for z=0 galaxies, suggesting an evolution of the G/D ratio with redshift as well as metallicity. By constraining these scaling relations at lower metallicities, we will be able to more accurately use observations to characterise the properties of these galaxies, such as their star-formation efficiencies and molecular gas content.

In this paper, we study the relationship between metallicity, gas, star-formation and dust content, in a sample of sub-solar metallicity galaxies in the cluster \Cl\ at z=1.99 \citep{ref:R.Gobat2011}. Despite residing in a cluster environment, these galaxies are consistent with the MZR at z=2 (as discussed in Section~\ref{sec:StackZ}) as well as the upper half of the main sequence (MS) of star-formation (Section~\ref{sec:sfrsSED}), and are in this sense representative of the `normal' population of galaxies for the purposes of this work. The study of these galaxies therefore allows us rare insight into the physics of the numerous sub-solar enrichment galaxies at this redshift, at masses lower than the characteristic mass of the stellar mass function at z=2 (log$_{10}$(M$^{*}$/M$_{\odot}$)$\sim$10.4, \citealt{ref:I.Davidzon2017}; see also \citealt{ref:O.Ilbert2010, ref:O.Ilbert2013, ref:A.Tomczak2014}). We target the CO[4-3] and CO[1-0] transitions in our galaxies, in addition to the dust continuum emission at 870$\micron$ and 2mm. We describe our observations and data reduction in Section~\ref{sec:method}, and present our stacked gas and dust constraints in Section~\ref{sec:Stacks}. We discuss the implications of our results in the context of star-formation rate L$^{\prime}_{\rm CO}$-SFR correlations, as well as scaling relations between \aco\ and G/D with metallicity in Section~\ref{sec:discussion}. We summarise in Section~\ref{sec:conc}. Throughout this paper we use a $\Lambda$CDM cosmology with H$_{0}$=70kms$^{-1}$Mpc$^{-1}$, $\Omega_{M}$=0.3 and $\Omega_{\Lambda}$=0.7. We adopt a Chabrier initial mass function (IMF, \citealt{ref:G.Chabrier2003}).

\section{Sample Selection and Main Data Sets}
\label{sec:method}

\begin{figure} 
\centering
 \includegraphics[width=0.47\textwidth]{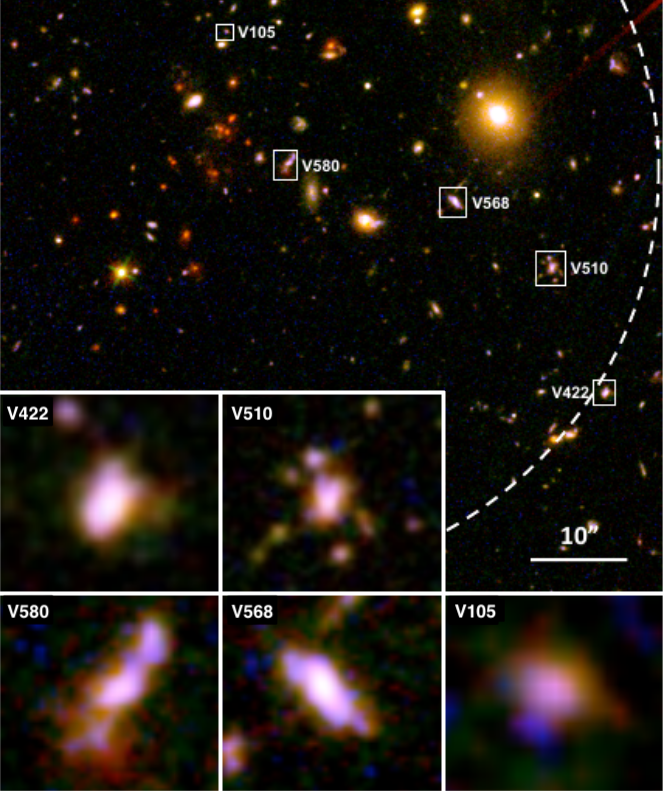}
    \caption{Colour image of the galaxy cluster \Cl, using HST/WFC3 filters F606W, F105W and F140W. The five low gas-phase metallicity galaxies discussed in this paper are indicated by white rectangles, and labelled with their IDs. The scale at the bottom right corner gives the angular scale of the main image, the cut-outs are zoomed versions of the individual galaxies as indicated. The white dashed circle shows the cluster R$_{200}$$\sim$0.4~Mpc physical radius \citep{ref:R.Gobat2013}.}
    \label{fig:ColourImage}
\end{figure}

The spectroscopically-confirmed galaxy population of \Cl\ has been studied at many different wavelengths to date, spanning X-ray to radio observations \citep{ref:R.Gobat2011, ref:R.Gobat2013, ref:V.Strazzullo2013, ref:V.Strazzullo2016, ref:V.Strazzullo2017, ref:F.Valentino2015, ref:F.Valentino2016, ref:R.Coogan2018a}. A subset of the $\sim$30 confirmed cluster members were selected for Subaru/Multi-object InfraRed Camera and Spectrograph (MOIRCS) near-infrared (IR) follow-up in \citet[][V15]{ref:F.Valentino2015}, based on their spectroscopic WFC3 redshifts \citep{ref:R.Gobat2013}, and their star-forming \textit{UVJ} colours \citep{ref:V.Strazzullo2013}. Seven of these targeted star-forming members were detected by MOIRCS, as were two galaxies known to host X-ray AGN \citep{ref:R.Gobat2013}. We exclude these AGN from our analyses, as the ionising radiation from the AGN is known to alter line ratios compared to pure star-forming galaxies (as demonstrated by e.g. the BPT diagram, \citealt{ref:L.Kewley2013}), rendering metallicity estimates unreliable. Here, we discuss the five of these seven star-forming galaxies for which we also have follow-up ALMA and VLA observations, allowing us to probe their ISM properties.

A colour image of the cluster is shown in Fig.~\ref{fig:ColourImage}, alongside cutouts of the cluster members discussed here. As shown in V15, these galaxies extend from the core of the cluster towards the outskirts, reaching $\sim$R$_{200}$ of \Cl\, at a physical radius of $\sim$0.4~Mpc. The cores of these galaxies do not appear highly dust obscured at rest-frame optical wavelengths, demonstrated by their blue colours in Fig.~\ref{fig:ColourImage}, implying ongoing star-formation activity.

To study the dust and molecular gas content of our targets, we use the ALMA and VLA data described in \citet{ref:R.Coogan2018a}: CO[4-3] line and 2mm continuum fluxes were derived from ALMA band 4 data, CO[1-0] line fluxes from VLA Ka-band observations and 870$\micron$ continuum measurements from ALMA band 7 images.

\begin{table*}
\begin{tabular}{ccccccccccc}
\hline
ID & RA & Dec & z$_{\rm opt}$ & 12 + log(O/H) & log$_{10}$(M$_{\star}$) & $\delta_{\rm MZR}$ & $ \delta_{\rm MS}$ & SFR$_{\rm H\alpha}$ & SFR$_{\rm UV}$  & E(B-V)$_{\rm neb}$\\
 & deg & deg & -  & PP04 & log$_{10}$$(M_{\odot}$) & dex & dex & $M_{\odot}yr^{-1}$ &$M_{\odot}yr^{-1}$ & mag\\
 \hline
V568 & 222.30242 & 8.93874 & 1.987$\pm$0.001 & $<$8.26 & 9.66$\pm$0.30 & $<$-0.03 & 0.52 & 39$\pm$5 & 41$\pm$8 & 0.15\\
V580 & 222.30702 & 8.93978 & 2.001$\pm$0.001 & $<$8.48 & 10.06$\pm$0.30 & $<$0.06 & 0.34 & 53$\pm$10 & 31$\pm$8 & 0.20\\
V422 & 222.29823 & 8.93351 & 1.988$\pm$0.001 & 8.14$\pm$0.12 & 9.76$\pm$0.30 & -0.18 & 1.1 & 170$\pm$23 & 71$\pm$13 & 0.37\\
V510 & 222.29972 & 8.93691 & 1.988$\pm$0.001 & $<$8.34 & 9.86$\pm$0.30 & $<$-0.01 & 0.49 & 52$\pm$8 & 57$\pm$10 & 0.35\\
V105 & 222.30873 & 8.94340 & 1.988$\pm$0.001  & $<$8.59 & 9.16$\pm$0.30 & $<$0.42 & 0.08 & 6$\pm$2 & $<$19 & 0.37\\
Sample Ave. & - & - & 1.9904 & 8.34$\pm$0.13 & 9.79$\pm$0.13 & 0.01 & 0.49 & 52$^{+19}_{-11}$ & 41$\pm$9 & - \\
\end{tabular}
\caption{Physical properties of the five low-metallicity cluster galaxies, and their average properties. MOIRCS rest-frame optical spectroscopic redshifts are from V15. [NII] was individually undetected in all galaxies except V422, so 2$\sigma$ upper limits on the oxygen abundance are shown, based on the N2 = log$_{10}$([NII]/H$\alpha$) diagnostic and the PP04 calibration. The characteristic metallicity of the sample was derived from a weighted stack, see Section~\ref{sec:StackZ} for details. $\delta_{\rm MS}$ and $\delta_{\rm MZR}$ are the logarithmic offsets from the MS of star-formation \citep{ref:M.Sargent2014} and the MZR \citep{ref:H.Zahid2014a}, respectively (a negative value indicates an offset below the relations). The characteristic mass of the sample is the average of the five individual stellar mass estimates. H$\alpha$ SFRs, UV SFRs, and the nebular reddening values (used to derive the dust-corrected H$\alpha$ SFRs given) are also reported. The UV SFRs were calculated using continuum reddening. 3$\sigma$ upper limits on UV SFR are shown. The characteristic H$\alpha$ and UV SFR values of the sample are the medians of the individual SFRs.}
\label{tab:GeneralProps}
\end{table*}

\subsection{MOIRCS observations}
Our five star-forming galaxies were selected as part of a MOIRCS near-IR spectroscopic follow-up, for which the observations and data reduction are described in detail in V15. Each galaxy was observed through a HK500 grism slit of width 0.7", and the data was reduced using the MCSMDP pipeline \citep{ref:T.Yoshikawa2010}. These observations primarily measured the  [NII]6548, [NII]6583, [OIII]5007 and \Halpha\ emission lines, in order to derive metallicities, precise spectroscopic redshifts, and \Halpha\ SFRs. For galaxy V105, not presented in V15 but discussed in this paper, the same approach as in V15 was taken for deriving these quantities. The results of these measurements are listed in Tables~\ref{tab:GeneralProps} and \ref{tab:SFRs}.

\subsection{CO spectroscopy and dust continuum measurements}
\label{sec:extraction}

The ALMA band 4 data used to measure CO[4-3] fluxes contain a deeper, core observation (continuum root-mean-squared (RMS) noise 8$\micro$Jy/beam), and a shallower pointing (continuum RMS 24$\micro$Jy/beam) of the outskirts of the cluster. The VLA Ka-band data used to measure CO[1-0] emission in the galaxies is a single pointing with continuum RMS 3$\micro$Jy/beam.

We extracted CO spectra at fixed positions, corresponding to the rest-frame optical positions of the galaxies shown in Fig.~\ref{fig:ColourImage}. Line fluxes were measured directly in the uv-visibility plane with the GILDAS\footnote{http://www.iram.fr/IRAMFR/GILDAS} task \texttt{uv\_fit}. Visibilities were collapsed over the expected CO linewidth, chosen to correspond to the velocity range containing 90\% of the \Halpha\ line flux, using the individual \Halpha\ linewidths listed in Table~\ref{tab:SFRs}. It should be noted that these listed 90\%-flux linewidths, W$_{90\%}$, are Full Width Zero Power (FWZP) widths, whereas the \Halpha\ linewidths are Gaussian Full Width Half Maxima (FWHM). The choice of line profile is not expected to have a significant effect on our line flux measurements, as the fluxes are measured directly in the collapsed visibility data over the given velocity width. Additionally, the line FWHM for detected CO[4-3] transitions of other members of \Cl\ were found to be consistent between a double and a single Gaussian profile in \citet{ref:R.Coogan2018a}. Due to the small optical sizes of the galaxies (see Fig.~\ref{fig:ColourImage}) compared with the size of the beams for the CO datasets (1.19"$\times$0.96" and 0.70"$\times$0.64" for CO[4-3] and CO[1-0] respectively), as well as the low SNR achieved by the individual CO spectra, all spectra were extracted assuming a point spread function (PSF) response (1$\sigma$ spectral noise over a 100\kms\ width was 10~mJy\kms\ and 4~mJy\kms\ for the CO[4-3] core pointing and CO[1-0] respectively). In order to avoid contamination from other galaxies, particularly in the cluster core, neighbouring galaxies were iteratively modelled with GILDAS and removed from the data, before measuring the properties of the low-metallicity galaxy population. The exception to this was galaxy V580, which was modelled simultaneously with its dusty companion (galaxy ID A2 in \citealt{ref:R.Coogan2018a}).

2mm continuum fluxes were measured in the uv-plane, after collapsing the data over the line-free spectral windows. The 870$\micron$ map is a mosaic formed of 7 single pointings, with a total area of 0.3~arcmin$^{2}$ centered on the cluster core. This therefore only covers the positions of galaxies V105, V580 and V568. However, we include this dataset in the analysis for these three galaxies (as well as the stacked sample) as it leads to a significant improvement in sensitivity of the dust emission maps ($\times$2 compared to the 2mm continuum). Continuum fluxes in the 870$\micron$ map were extracted using the software GALFIT \citep{ref:C.Peng2010}, using the full dirty beam on the dirty, calibrated images, as the mosaic format made the transportation into GILDAS format impractical. We have previously verified that continuum fluxes measured with Common Astronomy Software Applications \citep{ref:J.P.McMullin2007}, GILDAS and GALFIT for the dusty star-forming galaxies in this cluster are consistent \citep{ref:R.Coogan2018a}. All galaxies in our sample were again modelled as PSFs in the continuum (beams of 1.19"$\times$0.96" and 1.41"$\times$0.62" for 2mm and 870$\micron$ respectively).

\subsection{SED fitting and star formation rates}
\label{sec:sfrsSED}
We take stellar masses and SFRs for our sample as derived by \citet{ref:V.Strazzullo2016}, using Fitting and Assessment of Synthetic Templates \citep[FAST;][]{ref:M.Kriek2009} spectral energy distribution (SED) modelling. Continuum reddening values were calculated assuming a truncated star-formation history before correction to the higher nebular extinction, as is appropriate for the \Halpha-derived SFRs (e.g. \citealt{ref:D.Calzetti2000}, \citealt{ref:D.Kashino2013}, \citealt{ref:M.Pannella2014}, V15). The updated reddening values from \citet{ref:V.Strazzullo2016} resulted in reductions of the galaxy stellar masses and \Halpha\ SFRs by $\sim$0.4~dex and $\sim$0.2~dex respectively, compared with V15. The nebular E(B-V) values and H$\alpha$ SFRs are presented in Table~\ref{tab:GeneralProps}. These galaxies have $\delta_{\rm MS}$~=~SFR$_{\rm H\alpha}$/SFR$_{\rm MS}$ values between 0.08 and 1.1~dex, as shown in Table~\ref{tab:GeneralProps}. The median offset of our galaxies from the MS is 1.6$\sigma$, where all but one of the galaxies (V422) lie below the frequently adopted `cut-off' between the starburst and MS regimes, at 0.6~dex \citep{ref:G.Rodighiero2011, ref:M.Sargent2012}.

\section{Results}
\label{sec:Stacks}

In this section we describe our metallicity measurements, CO[1-0] and CO[4-3] line fluxes, dust continuum measurements and estimates for the molecular gas masses of our targets. CO fluxes are used to examine the relationship between CO luminosity and SFR in the low-metallicity regime. The combination of gas mass and L$^{\prime}_{\rm CO[1-0]}$ allows us to put constraints on the \aco\ conversion factor, and similarly, we can use the dust continuum measurements to constrain G/D ratios.

\subsection{Gas phase metallicity measurements}
\label{sec:StackZ}

We derive oxygen abundances with the N2 = log$_{10}$([NII]6583/\Halpha) diagnostic and the \citet[][PP04]{ref:M.Pettini2004} calibration using the MOIRCS spectra of our targets. Significant (SNR$>$5) \Halpha\ line detections were found for all five galaxies, but [NII] remained undetected with the exception of V422 where we find a weak detection (2.0$\sigma$). We therefore derived upper limits on the metallicities of the [NII]-undetected galaxies. Known disadvantages of the N2 calibration are that it saturates at high metallicity, as well as being dependent on both the ionisation parameter and the ratio of N/O \citep{ref:J.Baldwin1981, ref:KewleyDopita2002, ref:PerezMontero2009}. However, our galaxies are at significantly sub-solar metallicity, and the effect of the varying N/O ratio on the derived metallicities of these galaxies is negligible (V15). To further verify our N2-based metallicity values, we also use the O3N2 PP04 calibration to calculate metallicities of these galaxies, again using the MOIRCS data. We find that the O3N2 indicator gives systematically lower metallicity than N2, as has also been seen at high-z by several other studies (e.g. \citealt{ref:D.Erb2006}, \citealt{ref:C.Steidel2014}, \citealt{ref:H.Zahid2014a}). We find a median metallicity offset between the metallicity upper limits inferred with N2 and O3N2 respectively of 0.09 dex, within the systematic uncertainties of the two estimators (0.18~dex and 0.14~dex for N2 and O3N2 respectively, PP04). We use N2 as our primary metallicity estimator, allowing us to make direct comparisons with the relevant literature (e.g. \citealt{ref:G.Magdis2011, ref:G.Magdis2012}, V15).

We report the metallicities of the individual galaxies in Table~\ref{tab:GeneralProps}, as well as the offsets from the MZR at z$\sim$2. For the higher mass galaxies, with M$_{\star}>$10$^{9.6}$M$_{\odot}$, their metallicity constraints place them $\sim$0.02~dex below the expected value, based on the mass-metallicity relation at z=2 \citep{ref:H.Zahid2014a}. For V105, the galaxy with the lowest SFR and stellar mass, the upper limit on the oxygen abundance places it $\sim$0.4~dex above the MZR\footnote{Note that 4/5 of our galaxies were reported as lying below the z=2 MZR in V15. It is the reduction of the stellar masses, discussed in Section~\ref{sec:sfrsSED}, that brings these galaxies into agreement with the MZR (whilst metallicity values remain the same).}. The offsets $\delta_{\rm MZR}$ calculated with respect to the parametrisation of the MZR in \citet{ref:H.Zahid2014a} are consistent with those we derive for the MZR in both the N2 and O3N2 PP04 calibration, as reported by \citet{ref:C.Steidel2014}.

We derived a stacked metallicity estimate for all five galaxies combined. Following the same procedure as in V15, the continuum, \Halpha, [NII]6548 and [NII]6583 lines were modelled simultaneously in the stacked spectrum, using 1D Gaussian profiles. We imposed the positions of the [NII] lines to correspond to the redshift derived from the \Halpha\ line, as it has the highest SNR. We also imposed the standard constraint that the ratio of the line fluxes between the two [NII] lines must be equal to 3.05 \citep{ref:P.Storey2000}. The amplitude of a frequency-independent continuum contribution was left free to vary. We verified through a jackknifing analysis that our results are not heavily biased by individual galaxies, and that the method of stacking (optimal, median, average) does not significantly affect our results. The average metallicity and associated error we quote for the stacked galaxies is based on a Monte Carlo resampling, for which we created 1000 stacked spectra, each composed of a weighted average of individual, resampled galaxy spectra.

We find a weighted-average, stacked metallicity value Z=8.34$\pm$0.13. This error includes an additional factor derived from the spread of the metallicities from the jackknifing analysis, without the systematic uncertainty on the N2 PP04 calibration itself (0.18~dex). V15 used a sample of 6 low-metallicity cluster galaxies (4 of which are common to this study, with the exception of V105), to derive a mean metallicity value of Z=8.261$\pm$0.083, consistent with what we find for our sample. With an average mass log$_{10}$(M$_{\star}$/M$_{\odot}$)~=~9.79$\pm$0.30, this stacked metallicity measurement implies an offset of $\sim$0.01~dex above the \citet{ref:H.Zahid2014a} MZR (the mean offset $\delta_{\rm MZR}$ with respect to the \citet{ref:C.Steidel2014} MZR at z=2.3 is -0.03~dex, consistent within 0.2$\sigma$).

\begin{figure*}
\hspace{-1.8mm}
\begin{minipage}{178.5mm}
    \centering
    \includegraphics[width=\textwidth]{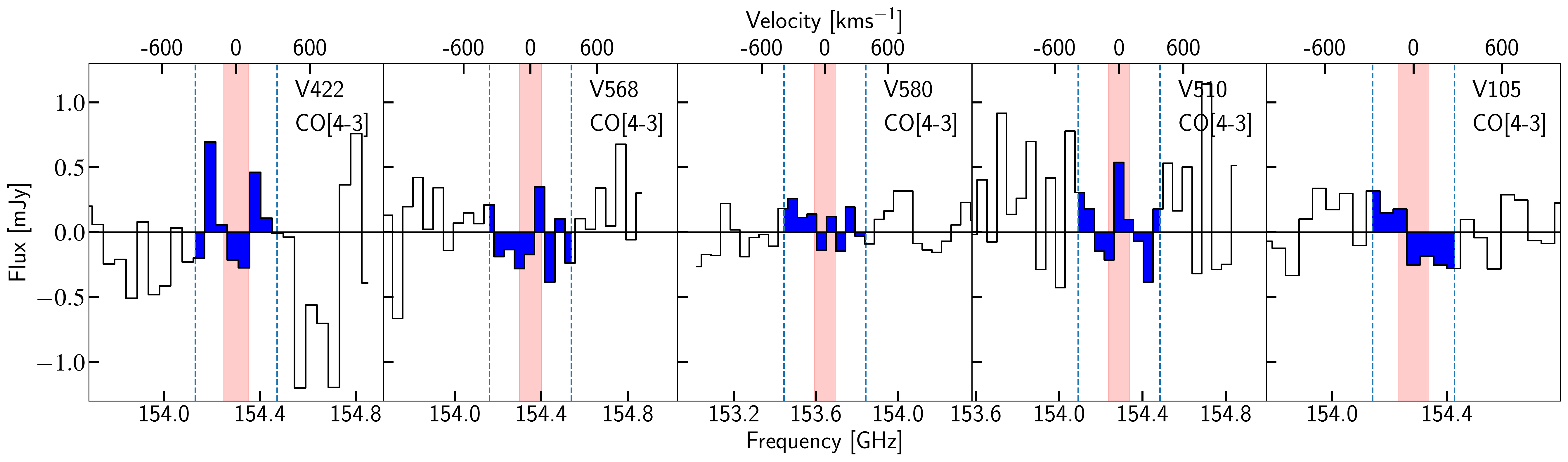}
\end{minipage}
\begin{minipage}{180mm}
\hspace{-1.9mm}
    \centering
    \includegraphics[width=\textwidth]{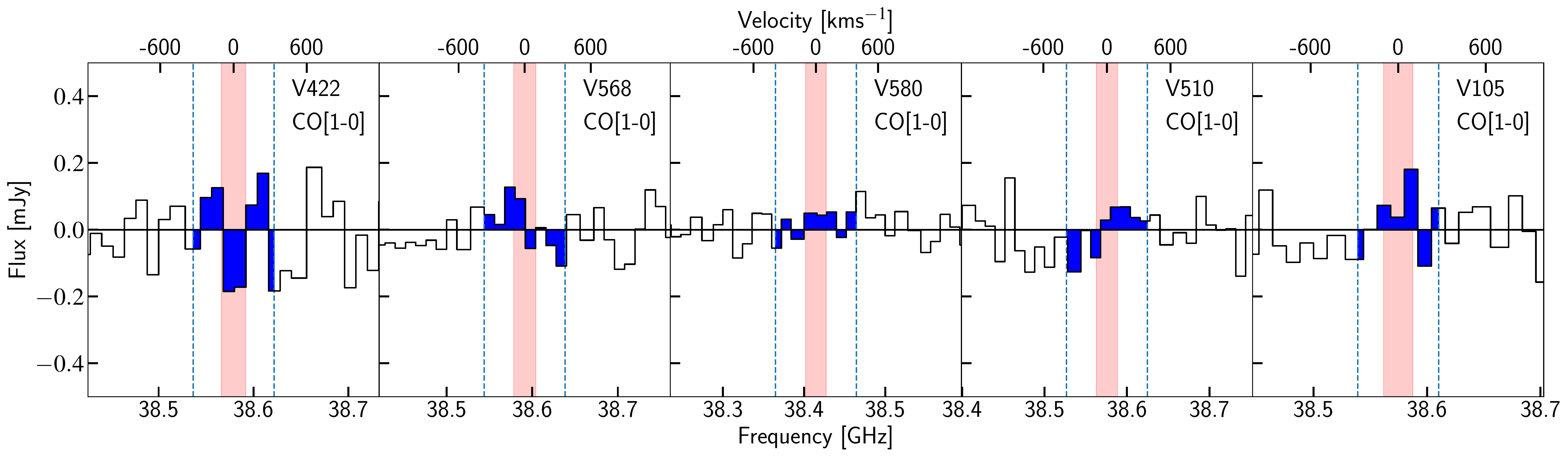}\hfill
    \caption{Top panel (lower panel): individual CO[4-3] (CO[1-0]) spectra binned into 90 km/s channels. The expected central position of the CO lines based on optical spectroscopic redshifts and associated errors are shown by the red vertical bands. The expected widths of the CO lines are shown by the blue shading of the spectra and the vertical dashed lines. The CO[4-3] and CO[1-0] transitions of all galaxies are individually undetected.}
    \label{fig:Lines}
 \end{minipage}
\end{figure*}

\begin{table}
\centering
\begin{tabular}{ccccc}
\hline
ID & I$_{\rm CO[4-3]}$ & I$_{\rm CO[1-0]}$ & F$_{\rm 870GHz}$ & F$_{\rm 2mm}$ \\
 & mJy \kms & mJy \kms  & $\micro$Jy & $\micro$Jy\\
\hline
V568 & $<$118 & $<$39 & $<$346 & $<$32 \\
V580 &  $<$74 & $<$38 & $<$203 & $<$22\\
V422 &  $<$221 & $<$79 & - & $<$57 \\
V510 &  $<$218 & $<$58 & - & $<$58 \\
V105 &  $<$89 & $<$31 & $<$239 & $<$30 \\
Sample Ave. &  $<$47 & $<$20 & - & $<$14 \\
\end{tabular}
\caption{Integrated line fluxes of the galaxies at the expected positions of the CO[4-3] and CO[1-0] lines, taken over the 90\%-flux widths shown in Table~\ref{tab:SFRs}, W$_{90\%}$. Continuum upper limits at 870$\micron$ and 2mm are also shown. 3$\sigma$ upper limits are given. A dashed entry for a galaxy indicates that the galaxy was not covered by the 870$\micron$ mosaic. We do not give a sample average 870$\micron$ flux, as not all galaxies are present in the 870$\micron$ data.}
\label{tab:ICO}
\end{table}

\begin{table*}
\begin{tabular}{ccccccccccc}
\hline
ID & log$_{10}$(M$_{mol}$, KS) & log$_{10}$(L$^{\prime}_{\rm CO[1-0]}$) & $\alpha_{\rm CO}$ & log$_{10}$(M$_{d}$)  & G/D & $\sigma_{H\alpha}$ & W$_{90\%}$ \\
 & log$_{10}$$(M_{\odot})$ & log$_{10}$($\rm K~kms^{-1}~pc^{2})$ & $\rm M_{\odot}(K~kms^{-1}~pc^{2}$)$^{-1}$ & log$_{10}$$(M_{\odot})$ & - & \kms & \kms \\
 \hline
V568 & 10.50$^{+0.08}_{-0.10}$ & $<$9.88 & $>$4.3 & $<$8.26 &  $>$176 & 224$\pm$14 & 734$\pm$45\\
V580 & 10.62$^{+0.10}_{-0.13}$ & $<$9.88 & $>$5.5 & $<$8.04 &  $>$375 & 202$\pm$27 & 779$\pm$29\\
V422 & 11.03$^{+0.08}_{-0.10}$ & $<$10.19 & $>$6.9 & $<$8.76 &  $>$183 & 237$\pm$9 & 663$\pm$89\\
V510 & 10.61$^{+0.09}_{-0.11}$ & $<$10.05 & $>$3.6 & $<$8.77 &  $>$69 & 233$\pm$20 & 763$\pm$66\\
V105 & 9.84$^{+0.16}_{-0.25}$  & $<$9.78 & $>$1.1 & $<$7.98 &  $>$51 & 169$\pm$40 & 554$\pm$130\\
Sample Ave. & 10.61$^{+0.14}_{-0.20}$ & $<$9.58 & $>$10.61 & $<$7.92 & $>$478 & - & 753$\pm$22\\
\end{tabular}
\caption{ISM properties of individual galaxies in our sample, and of a stack of all 5 objects. log$_{10}$(M$_{mol}$, KS) is the molecular gas mass inferred from SFR$_{\rm H\alpha}$ via the integrated Kennicutt-Schmidt relation for main-sequence galaxies \citep{ref:M.Sargent2014}. CO[1-0] line luminosities, derived $\alpha_{\rm CO}$ conversion factors, dust masses, total G/D ratios, \Halpha\ linewidths from V15 and the expected CO 90\%-flux widths (W$_{90\%}$) derived from these \Halpha\ lines are also listed. Upper limits on dust mass include the systematic errors discussed in the text. All upper/lower limits are placed at 3$\sigma$.}
\label{tab:SFRs}
\end{table*}

\subsection{Molecular line emission: CO[1-0] and CO[4-3]}
\label{sec:CO}
\begin{figure*}
\begin{minipage}{177mm}
\centering
    \includegraphics[clip, width=0.3333333\textwidth, trim=0cm 0.2cm 0cm 0.2cm]{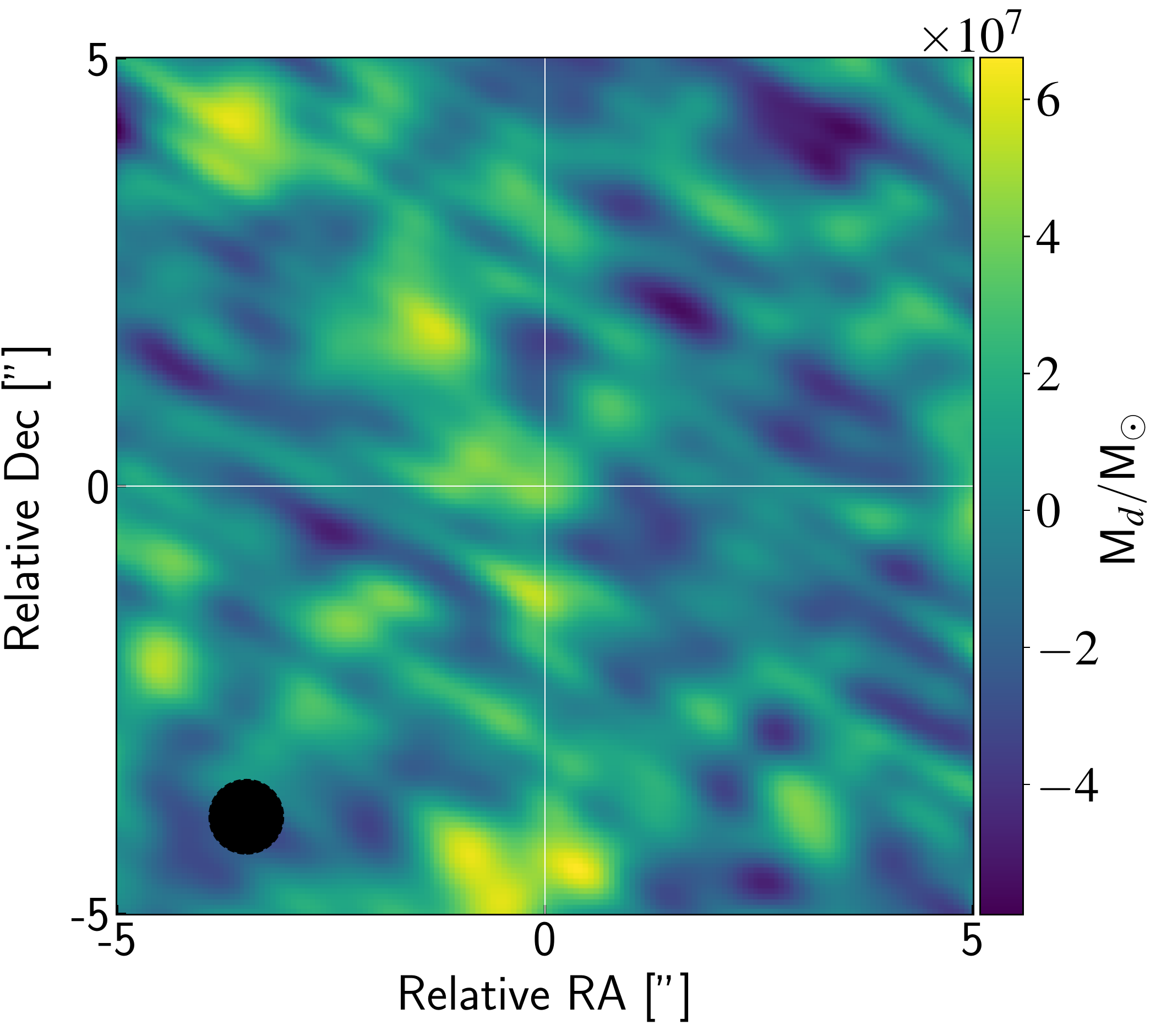}\hfill
    \includegraphics[clip, width=0.326\textwidth, trim=0cm 0.24cm 0cm 0cm]{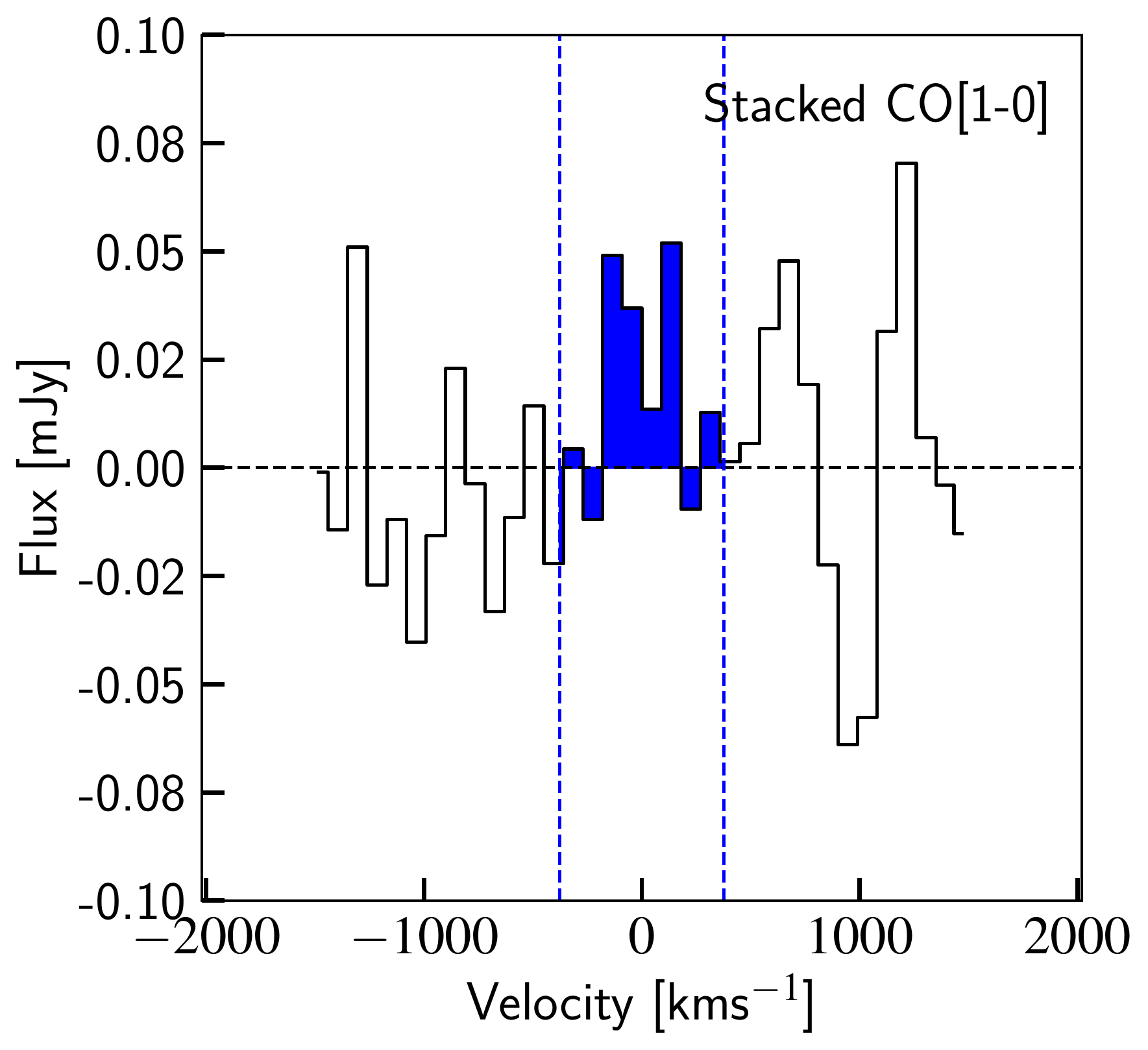}\hfill
    \includegraphics[clip, width=0.325\textwidth, trim=0cm 0.24cm 0cm 0.24cm]{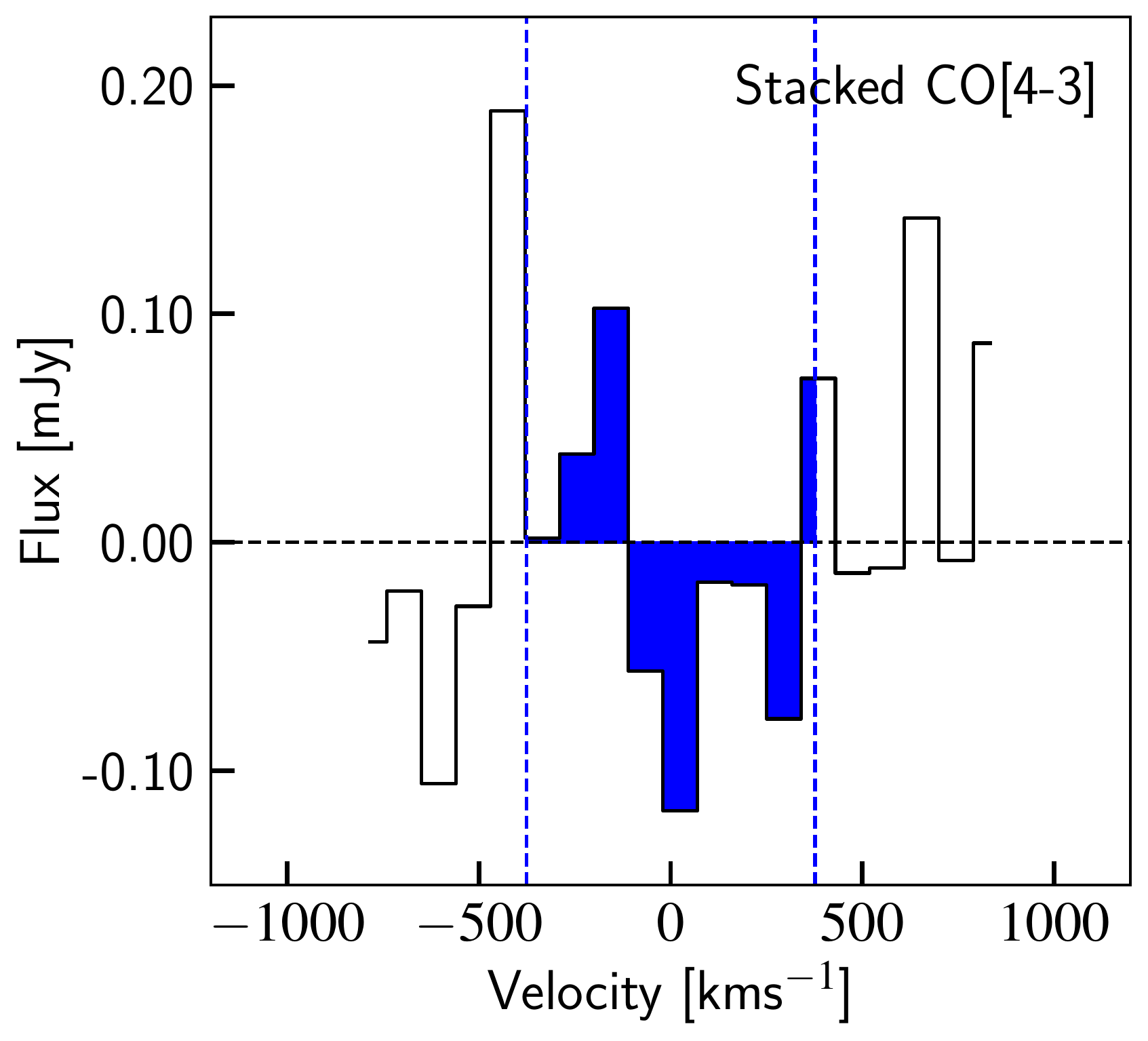}\hfill
    \caption{Left: stacked dust mass image for the five galaxies, created by combining the dust mass images from the 2mm data and 870$\micron$ data, as described in the text. The image is centered on the position of the stacked galaxies, and the white crosshair highlights the position of the peak pixel, used to derive the upper limit on dust mass. The average of the circularised beams for the two continuum datasets is shown in the bottom left corner. Center: weighted-average stacked CO[1-0] spectrum of the five galaxies, binned into 90 km/s channels and shown against velocity relative to the expected position of the CO[1-0] line. The blue shading illustrates the average linewidth (containing 90\% of the flux) of the galaxies, W$_{90\%}$=753$\pm$22\kms, over which the stacked measurement was taken. Right: as for the center panel, but for the CO[4-3] spectral stack. The CO[4-3] spectrum covers a smaller velocity range than the CO[1-0] stack, as the spectral resolution was coarser in the sidebands used for continuum measurements.}
    \label{fig:duststackImage}
   \end{minipage}
\end{figure*}

As discussed in Section \ref{sec:method}, by using the galaxy optical redshifts and the \Halpha\ linewidths, we can extract CO spectra and line fluxes for each galaxy over the expected velocity range. The CO[4-3] spectra for the individual galaxies are displayed in the top panel of Fig.~\ref{fig:Lines} and show no significant line detections. Similarly, the galaxies in our sample also remain individually undetected in the CO[1-0] transition (Fig.~\ref{fig:Lines}, lower panel). Upper limits for all line fluxes are presented in Table~\ref{tab:ICO}. We perform a weighted mean stack of all 5 galaxies (Fig.~\ref{fig:duststackImage}), and calculate the stacked line fluxes over the mean FWZP linewidth of the individual galaxies, $\Delta$v=753$\pm$22\kms. This corresponds to the weighted mean linewidth ($\langle\sigma_{H\alpha}\rangle$~=~231\kms), transformed to a FWZP containing 90\% of the \Halpha\ flux. The error on this average linewidth reflects the error on the weighted mean. We perform Monte Carlo Markov Chain (MCMC) simulations to take into account the statistical errors on the redshifts of the individual spectra, displacing the individual spectra before stacking, according to the redshift sampled from a Gaussian distribution. The effect of this on the output stacked line fluxes is negligible. The flux uncertainties on the stacked spectra have been calculated using standard error propagation on the weighted mean of the individual channels. We verify that the error on the flux arising from the error on the linewidth is negligible, using a resampling approach over a Gaussian linewidth distribution. It should also be noted that the upper limits on these CO line fluxes are proportional to the square root of the linewidth. A factor of two difference in linewidth, for example, would therefore alter the derived (logarithmic) CO luminosities by $\pm$0.15~dex, and would not affect the conclusions of this work.

We find no detection of CO[4-3] emission, even in this stacked data, where the noise on the stacked spectrum is reduced by a factor $\sim$$\sqrt{5}$ compared to the individual spectra. We therefore place a 3$\sigma$ upper limit on the stacked CO[4-3] line flux of $<$47mJy~\kms, shown in Table~\ref{tab:ICO}. In the stacked CO[1-0] spectrum, we find a 3$\sigma$ upper limit of $<$20mJy~\kms. In both cases, we have scaled the errors on the spectra using the prescription that \texttt{Var}(S$_{\rm \nu}$/$\sigma_{\rm \nu}$)~=~1 (where S$_{\rm \nu}$ and $\sigma_{\rm \nu}$ are the flux and associated error respectively in each line-free channel), thereby ensuring that they represent a true Gaussian distribution and are not underestimated. At the mean redshift of our targets, this corresponds to to an upper limit log(L$^{\prime}_{\rm CO[1-0]})<$9.58 (see Table~\ref{tab:SFRs}). We discuss the potential implications of these low CO luminosities in Section~\ref{sec:lco}.

\subsection{Dust continuum emission}
\label{sec:duststack}

In order to estimate the dust mass of the cluster galaxies, we convert between submillimetre continuum flux and dust mass, $M_{d}$. This can be done separately for both the 870$\micron$ and 2mm continuum fluxes. As we cannot constrain the temperature of the dust with these data, we assume the shape of the z=2 IR-to-sub-mm MS SED from \citet{ref:G.Magdis2012}, the shape of which is consistent with the individual limits on our continuum fluxes. We then calculate dust mass using:

\begin{equation}
M_{d}=\frac{L_{IR}}{125\langle U \rangle}\propto\frac{4\pi d_{L}^{2}}{125 \langle U \rangle} \times F_{\lambda}
\end{equation}

where d$_{L}$ is the luminosity distance of the galaxy, F$_{\lambda}$ is either the 870$\micron$ or 2mm flux density and $\langle$U$\rangle$ is the mean, dust-weighted intensity of the stellar radiation field. The proportionality reflects the variable bolometric correction used to convert from sub-mm flux to L$_{\rm IR}$ for a given template shape (determined by our assumed value of $\langle$U$\rangle$=22, in accordance with the redshift-evolution of this parameter measured in \citealt{ref:M.Bethermin2015}). 125 is the average scaling factor between L$_{\rm IR}$/$M_{d}$ from \citet{ref:G.Magdis2012}. If the dust temperatures in our galaxies were warmer than the template, as observed locally in low-mass and low-metallicity dwarf galaxies (e.g. \citealt{ref:F.Walter2007}), this would in turn decrease the derived dust mass, and increase the G/D ratios. As we do not constrain the temperature, we allow a systematic error of 0.15~dex for these dust masses \citep{ref:DraineLi2007}. In this way, we first calculate separate dust mass estimates from the 870$\micron$ and 2mm measurements (for V105, V580 and V568). The final dust mass of each galaxy is then calculated by combining the two dust mass estimates, weighted according to their errors. Where appropriate, the upper limit on the dust mass is calculated from this weighted average and associated error. For V422 and V510, the dust mass comes only from the 2mm measurements.

Similarly to the CO emission, we find no significant dust continuum detections for these galaxies individually. The constraints on these dust masses are given in Table~\ref{tab:SFRs}. We therefore created a stacked dust map for the five low-metallicity galaxies. To do this, we first made individual dust mass maps for each galaxy. We created 10"$\times$10" cutouts centered on each galaxy, in both the 870$\micron$ and 2mm data. This was done using the dirty image maps, having removed all other sources and corresponding sidelobes by simultaneous flux modelling of the entire field in GALFIT, using the full known dirty beam pattern. Each flux emission map (2mm or 870$\micron$) was then converted to a dust mass map, pixel by pixel, for each galaxy, by converting between flux and dust mass as described above. All of these individual dust maps were then stacked in the image plane, by performing a weighted co-add of the images at each pixel position. The weights were determined by the primary beam corrected RMS noise in the individual image cutouts. The stacked dust map is shown in Fig.~\ref{fig:duststackImage}.

As this image combines two datasets imaged with different synthesised beam shapes, PSF-fitting with GALFIT was not practical. If - as motivated in Section~\ref{sec:extraction} based on their small optical sizes - the galaxies in the stack are unresolved, we can measure the dust mass associated with the stack by reading off the value of the central pixel in the stacked image and interpreting it as a peak flux measurement. We note that, as the circularised FWHM of the synthesized beams differ by only $\sim$10\%, the two datasets sample similar spatial scales and have been imaged with the same pixel size. We find a 3$\sigma$ upper limit on the stacked dust mass of M$_{d}<$7.4$\times$10$^{7}$~M$_{\odot}$. This small amount of dust seems consistent with initial expectations for low-metallicity galaxies, and we discuss how the dust and gas in these galaxies scale with metallicity in Section~\ref{sec:GD_Z}.

\subsection{Molecular gas masses}
\label{sec:gasmass}
We have calculated the molecular gas content of the galaxies by taking the H$\alpha$ SFR of each galaxy, and using the integrated Kennicutt-Schmidt (KS, equation 5 of \citealt{ref:M.Sargent2014}) law to convert between SFR and molecular gas mass M$_{mol}$ as below:

\begin{equation}
M_{mol} = SFR \times \frac{1}{SFE}
\end{equation}

\begin{figure*}
    \centering
    \begin{minipage}{175mm}
    \includegraphics[width=0.5\textwidth]{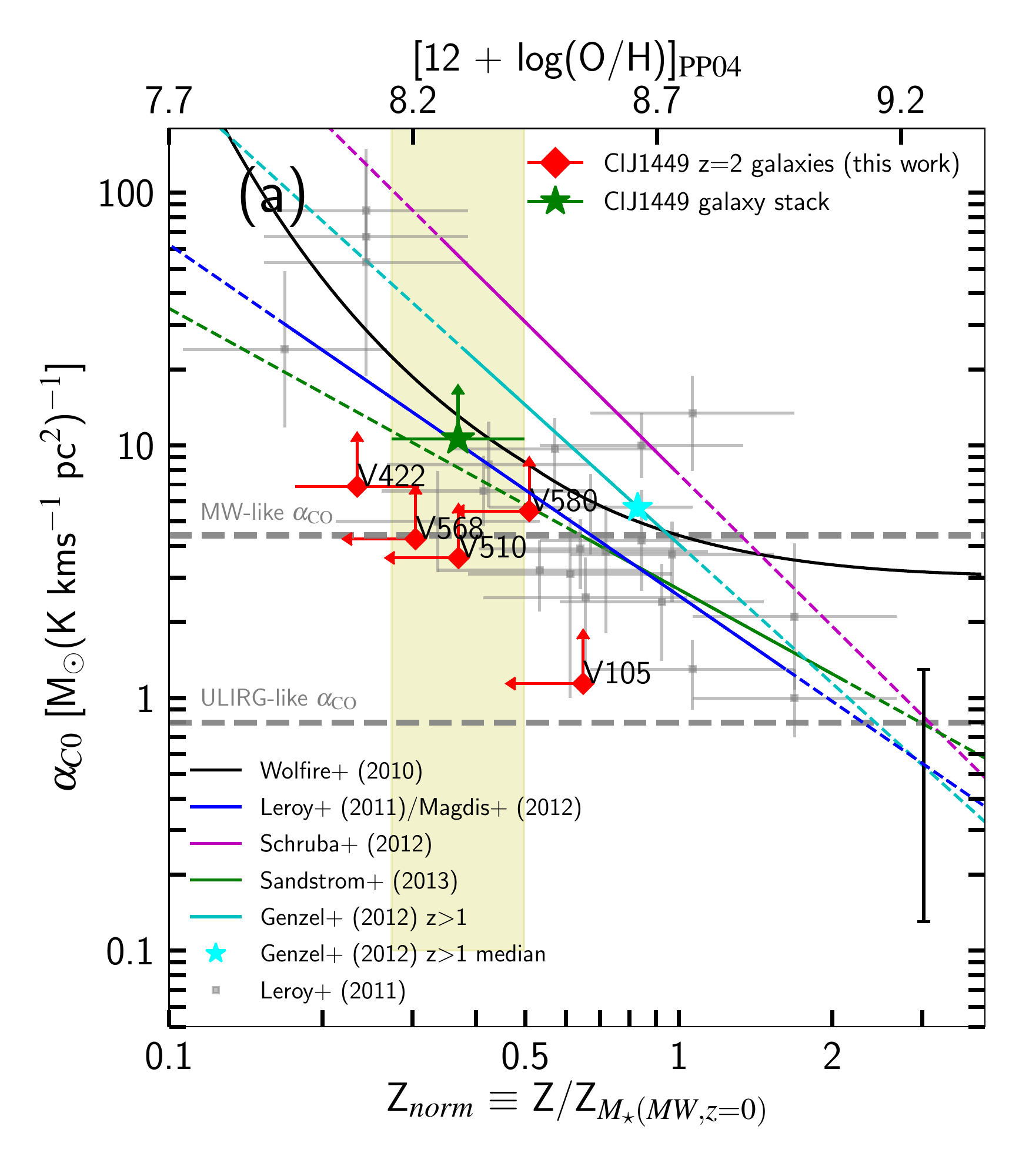}
    \includegraphics[width=0.5\textwidth]{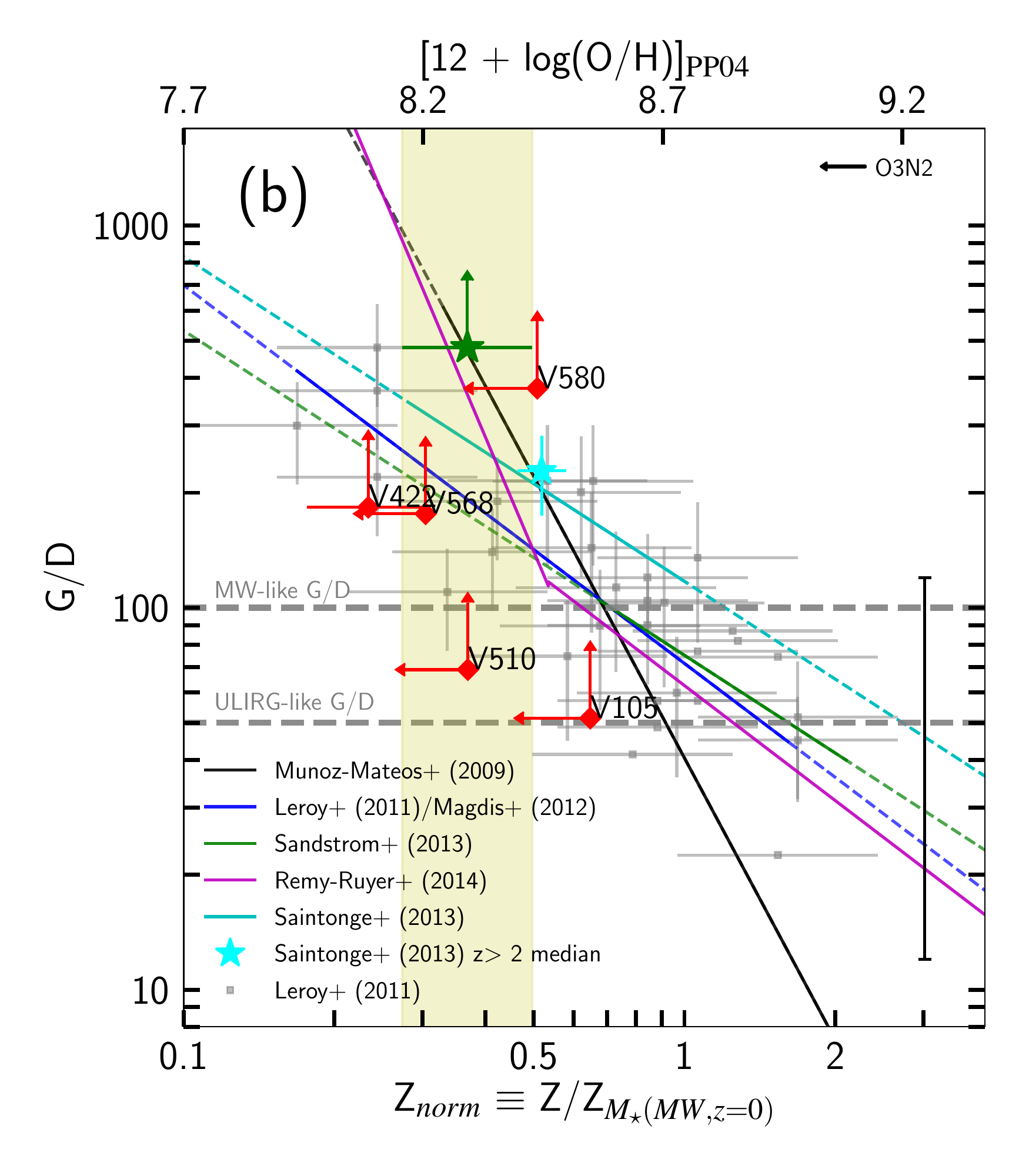}
    \caption{Left: CO-to-\Htwo\ conversion factor $\alpha_{\rm CO}$ as a function of metallicity. 3$\sigma$ lower limits on $\alpha_{\rm CO}$ are shown. Typical $\alpha_{\rm CO}$ values for MW- and ULIRG-like galaxies are $\alpha_{\rm CO}$=4.4 and $\alpha_{\rm CO}$=0.8 respectively \citep{ref:Downes1998}, shown by the grey dashed lines. Right: Gas-to-dust ratio as a function of metallicity. 3$\sigma$ lower limits on the G/D ratio are shown. Typical G/D ratios of 100 and 50 for MW- and ULIRG-like galaxies respectively are indicated by the grey dashed lines \citep{ref:P.Santini2014}. In both panels, the comparison galaxy sample shown in grey comprises local galaxies, z$\sim$0.5-2.5 galaxies, local ULIRGS and submillimetre galaxies, taken from \citet{ref:A.Leroy2011a} and \citet{ref:G.Magdis2012}. The upper x-axis gives the equivalent PP04 metallicity, valid for the individual galaxy data points reported in this figure. The horizontal black arrow in the top right corner of panel (b) indicates the median offset in metallicity that would result if the O3N2, rather than the N2 diagnostic were used. Previous literature trends are indicated in the legends, where for observational studies solid lines indicate the metallicity range sampled by the corresponding data. (The dashed line segments illustrate the extrapolation of these trends to higher/lower metallicity). Each galaxy from our sample is is shown by a red diamond. We show a representative error bar of 1~dex in the lower right corner, to illustrate by how much the quantity on the y-axis would decrease if the SFE was 10$\times$ higher, i.e. gas masses 10$\times$ lower. The stacked limit for the 5 galaxies is shown as a green star. The yellow shaded region shows the error on the mean (stacked) metallicity of the galaxies.}
    \label{fig:alphaCO_Z}
   \end{minipage}
\end{figure*}

As we do not know the star-formation efficiencies (SFEs) of the galaxies, we assume that SFE increases smoothly with distance above the MS (e.g. \citealt{ref:M.Sargent2014, ref:L.Tacconi2018}), for each individual galaxy, to obtain molecular gas masses from SFR$_{\rm H\alpha}$. The assumption of these SFEs (corresponding to gas depletion times between $\sim$0.6-1.1~Gyr for our galaxies, with a sample median of 0.8~Gyr) is motivated by the position of our targets relative to the main-sequence of star-formation, and the consistency between the UV and \Halpha\ SFRs discussed below. We have no evidence that the galaxies have significantly increased SFEs or high gas excitation ratios. However, we also indicate the effect of a strongly increased SFE in Fig.~\ref{fig:alphaCO_Z}. The molecular gas masses derived assuming a MS SFE are shown in Table~\ref{tab:SFRs}. It is these gas masses that are used in the remainder of the analysis.

\section{The impact of a low gas-phase metallicity on ISM scaling relations at z$\sim$2}
\label{sec:discussion}

A key outcome of this study is the notable lack of gas and dust detections in our star-forming galaxies. If we take the \Halpha\ SFRs of our galaxies, we would expect to detect the CO[4-3] and CO[1-0] emission at $\geq$3$\sigma$ from 4/5 of our galaxies, assuming that the galaxies had BzK-like Spectral Line Energy Distributions (SLEDs, where BzK galaxies represent the `normal', solar-enrichment population at this redshift). In the continuum data, we would expect to detect 3/5 of our galaxies at $\geq$3$\sigma$ in the 2mm data (assuming a BzK-like SED), and to detect 2/3 of the galaxies covered by the 870$\mu$m mosaic. However, we see in Section~\ref{sec:Stacks} that we do not detect any of our galaxies in these datasets. We now discuss the implications of these low detection fractions.

\begin{figure*}
\begin{minipage}{175mm}
    \includegraphics[width=0.5\textwidth]{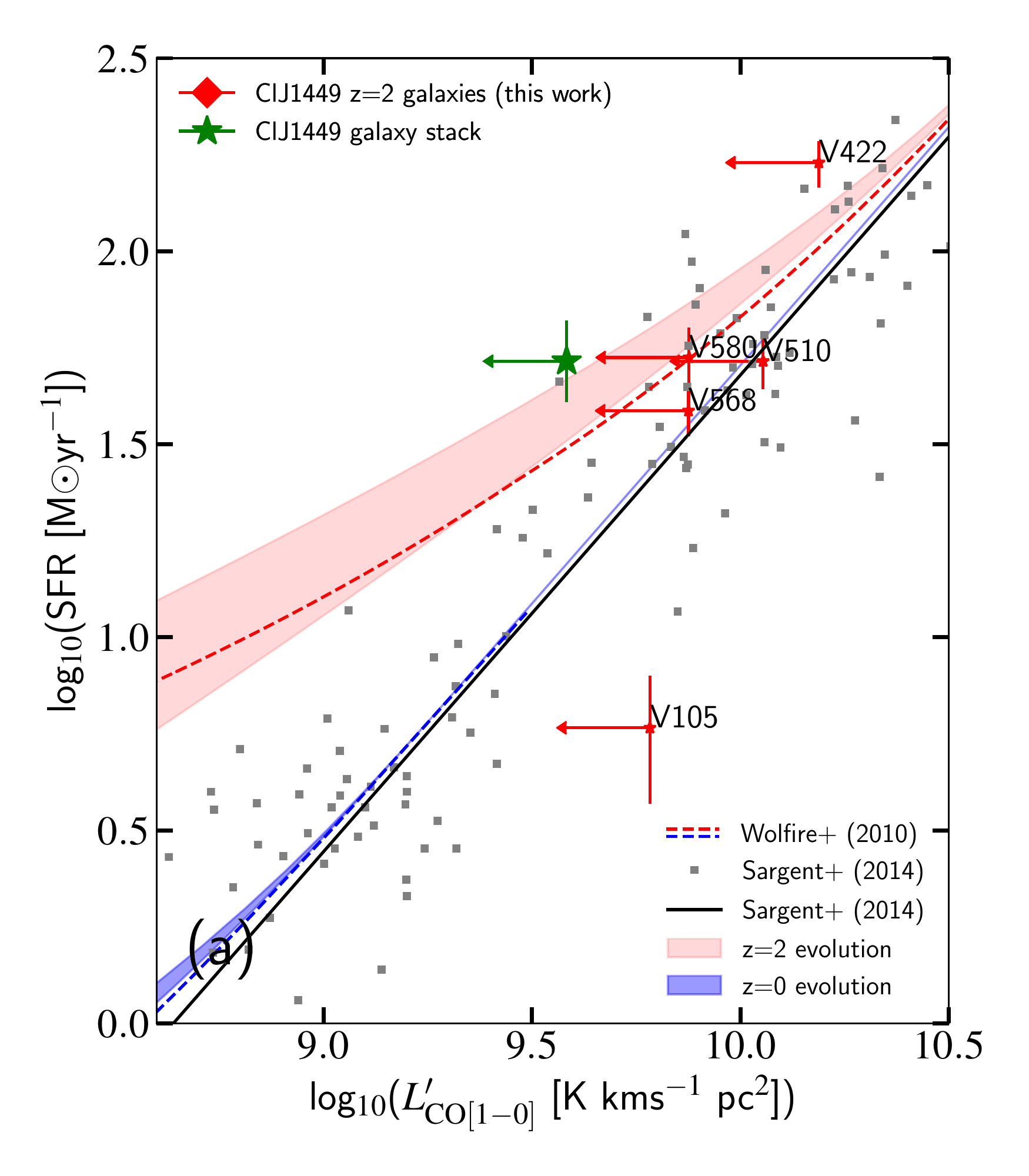}
    \includegraphics[width=0.5\textwidth]{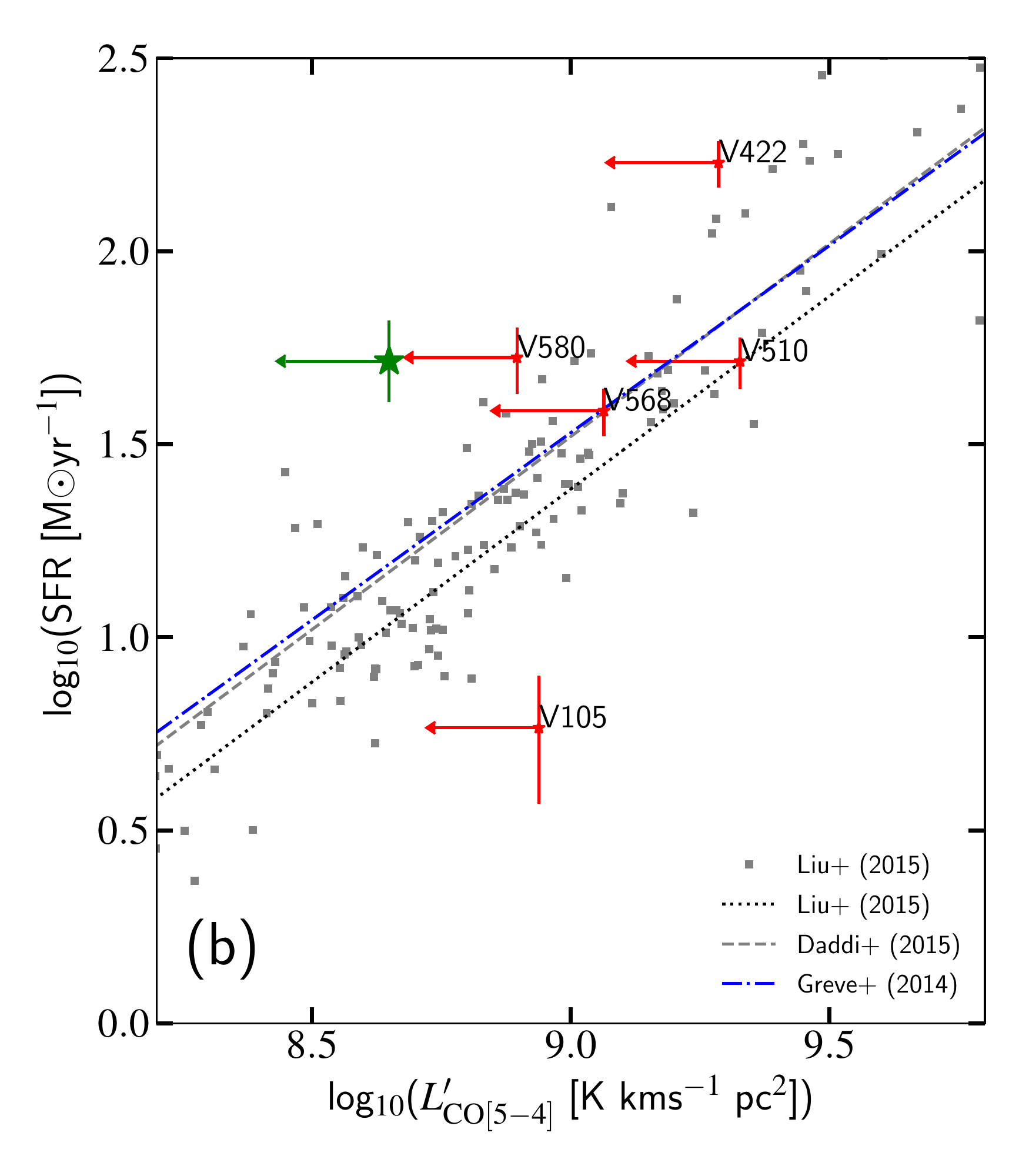}
    \caption{Left: SFR$_{\rm H_{\alpha}}$ vs. L$^{\prime}_{\rm CO[1-0]}$, where upper limits are located at 3$\sigma$. Predicted evolutionary trends are shown for z=0.08 and z=2 by the blue and red shaded curves, assuming an \aco\ vs. metallicity slope between -1 (lower edge) and -2 (upper edge), as well as evolutionary trends predicted by \citet{ref:M.Wolfire2010} (dashed lines). Right: SFR$_{\rm H_{\alpha}}$ as a function of L$^{\prime}_{\rm CO[5-4]}$. The green stars show the stacked upper limit on L$^{\prime}_{\rm CO[1-0]}$ (left) and L$^{\prime}_{\rm CO[5-4]}$ (right, from the stacked CO[4-3] data) respectively. The vertical green error bar on the stack shows the formal error on the median \Halpha\ SFR. Previous literature relations are indicated in the legend. We see evidence for metallicity-driven, systematically decreased line emission from the diffuse, extended (traced by CO[1-0]) and denser molecular gas phases (traced by CO[5-4]) in our targets.}
    \label{fig:SFR_SFR}
\end{minipage}
\end{figure*}

\subsection{SFR-L$^{\prime}_{\rm CO}$ correlations for z=2, low-metallicity galaxies}
\label{sec:lco}
Fig.~\ref{fig:SFR_SFR} (a) shows the relation between SFR and L$^{\prime}_{\rm CO[1-0]}$ for galaxies in our sample, together with literature data compiled by \citet{ref:M.Sargent2014} and containing MS galaxies at 0$<$z$<$3. With the exception of V105, we see that the 3$\sigma$ L$^{\prime}_{\rm CO[1-0]}$ limits for our galaxies lie on, or to the left of, the reference trend from \citet{ref:M.Sargent2014}. The stacked L$^{\prime}_{\rm CO[1-0]}$ limit is offset by $\sim$0.5~dex from this relation. This suggests that on average the galaxies have high SFRs given their L$^{\prime}_{\rm CO[1-0]}$, and the stacked limit occupies the same part of parameter space as high-SFE starburst galaxies.  However, the $\delta$MS values given in Table~\ref{tab:GeneralProps} indicate that our galaxies do not lie in the starburst regime in terms of their specific star-formation rates, with the exception of V422. An interpretation of this offset is then that the lack of L$^{\prime}_{\rm CO[1-0]}$ in the galaxies is driven by their low gas-phase metallicities, consistent with an increased amount of photo-dissociation of CO molecules in diffuse clouds.

We also show in Fig.~\ref{fig:SFR_SFR} (a) the predicted locus of `typical' z$\sim$0 and z$\sim$2 galaxies in SFR-L$^{\prime}_{\rm CO[1-0]}$ space. To trace this locus we start from the log-linear integrated Kennicutt-Schmidt relation - expressed in terms of SFR and M$_{\rm H2}$ - from \citet{ref:M.Sargent2014}. We then associate to a given SFR and M$_{\rm H2}$ a metallicity-dependent CO-to-H2 conversion factor \aco\ via the following three steps. We assume that SFR and M$_{\star}$ are linked via the main-sequence relation, and M$_{\star}$ and metallicity via the MZR. We then consider \aco\ vs. metallicity trends with a power law form Z$^{-\gamma}$, where 1$<\gamma<$2 (e.g. \citealt{ref:F.Israel1997, ref:S.Glover2011, ref:A.Leroy2011a, ref:D.Narayanan2012, ref:A.Schruba2012, ref:R.Amorin2016}), as well as the prescription from \citet{ref:M.Wolfire2010} (see also our Fig.~\ref{fig:alphaCO_Z} (a)). We have normalised the relationship between \aco\ and metallicity so that a galaxy with the stellar mass of the Milky Way (M$_{\star}$=5.7$\times$10$^{10}$M$_{\odot}$, \citealt{ref:T.Licquia2015}) has \aco=4.0\footnote{We omit the units of \aco\ for brevity, but give \aco\ in units of M$_{\odot}$(K~kms$^{-1}$~pc$^{2}$)$^{-1}$ throughout this paper.}. The expected SFR-L$^{\prime}_{\rm CO[1-0]}$ relation starts to curve away from a simple log-linear trend at low L$^{\prime}_{\rm CO[1-0]}$, with a `detachment point' that shifts to higher luminosities with increasing redshift. While V422 and V580 are offset from the log-linear SFR-L$^{\prime}_{\rm CO[1-0]}$ relation given by \citet{ref:M.Sargent2014},  the individual and stacked L$^{\prime}_{\rm CO[1-0]}$ limits are more closely consistent with the expected locus for z$\sim$2 galaxies. It should be noted here that the placement of starburst galaxies on this plane, offset from the SFR-L$^{\prime}_{\rm CO[1-0]}$ relation, is driven by a high SFE compared with MS galaxies, rather than by low-metallicity. The oxygen abundances of such high-z starbursts are typically higher than the galaxies discussed in this paper, consistent with solar metallicity (e.g. \citealt{ref:D.Rigopoulou2018}).

For further evidence that a CO-deficit rather than a SFE-enhancement is responsible for the position of our low-metallicity galaxies in SFR-L$^{\prime}_{\rm CO}$ space, we can also look at the SFR - L$^{\prime}_{\rm CO[5-4]}$ plane shown in Fig.~\ref{fig:SFR_SFR} (b). L$^{\prime}_{\rm CO[5-4]}$ was derived from the CO[4-3] data, assuming a MS, BzK-like excitation (I$_{\rm CO[5-4]}$~=~1.04$\times$I$_{\rm CO[4-3]}$, \citealt{ref:E.Daddi2015}). The assumption of a BzK- or starburst-like excitation between CO[4-3] and CO[5-4] does not significantly affect the derivation of this quantity. The comparison dataset in this figure comprises local ULIRGs and spirals \citep{ref:D.Liu2015}, BzK galaxies at z$\sim$1.5 \citep{ref:E.Daddi2015} and high-z SMGs \citep{ref:C.Carilli2013}. The individual limits on our galaxies again lie to the left of the literature relations, and the stacked L$^{\prime}_{\rm CO[5-4]}$ constraint lies at $\sim$0.5-0.7~dex below the L$^{\prime}_{\rm CO[5-4]}$ that would be expected from a galaxy with this SFR. This is particularly interesting, as \citet{ref:E.Daddi2015} have shown that there is a tight, linear correlation between the CO[5-4] luminosity of a galaxy and its IR luminosity (and therefore SFR), regardless of whether the galaxy is MS or starbursting, with increased scatter only for the most extreme objects across cosmic time \citep{ref:D.Liu2015}. The offset we see in Fig.~\ref{fig:SFR_SFR} (b) must therefore be a consequence of another physical property of the galaxies.

To test whether this offset could be due to a SFR timescale effect, rather than low-metallicity, we compare the \Halpha- and UV- SFRs in Table~\ref{tab:GeneralProps}. \Halpha\ is a short timescale ($<$10~Myr) tracer of star-formation, whereas UV emission traces star formation on timescales of $\sim$100~Myr \citep{ref:D.Calzetti2008, ref:R.Kennicutt2012}. The \Halpha- and UV- SFR estimates do not show large discrepancies, and are consistent to within $<$2$\sigma$, with the exception of very bright V422. The tension between these SFR indicators for V422 could indicate a recent increase in the SFR, but it should be noted that this comparison is dependent on the prescriptions of E(B-V) used for continuum and nebular extinction. This indicates that the SFRs of the galaxies have remained fairly constant for the last 100~Myr, i.e. show no indication of a rapidly declining SFR (which in turn would lead to low CO[5-4] emission compared to the SFR-L$^{\prime}_{\rm CO[5-4]}$ trend, as CO[5-4] flux effectively represents an instantaneous SFR-tracer following \citealt{ref:E.Daddi2015}). The clear offset of the stacked data in both panels of Fig.~\ref{fig:SFR_SFR} is therefore unlikely to be driven by timescale effects, and reinforces the conclusion that it is the low gas-phase metallicity in these galaxies that is driving the lack of CO emission at J=1-0 and J=4-3.

It is not clear to what extent one would expect the high-J transitions of CO to suffer the same effects of low enrichment as the low-J transitions. Studies of dwarf galaxies in the local Universe have revealed CO[3-2]/CO[1-0] and CO[2-1]/CO[1-0] line ratios consistent with those observed in local spiral galaxies (e.g. \citealt{ref:L.Sage1992, ref:D.Meier2001, ref:D.Cormier2014}), indicating that photo-dissociation impacts the CO Spectral Line Energy Distribution up to at least the J=3-2 transition in local low-metallicity galaxies. It is qualitatively understood that the effect of photo-dissociation is less prevalent in small regions of `dense' CO gas (e.g. \citealt{ref:A.Bolatto2008}), but it is not widely discussed to which CO excitation this extends to, and whether this is dependent on the individual galaxy. Aside from the destruction of CO by photo-dissociation, it is plausible that high-J excitation could be somewhat enhanced in low metallicity environments, due to the stronger radiation field. We see in our galaxies that, despite this possibly enhancing effect, the molecular ISM at densities of $\sim$10$^{5}$~cm$^{-3}$ ($\sim$40$\times$ the critical density of the CO[1-0] molecule) nevertheless displays a significant CO-deficit, compared to the tight SFR-L$^{\prime}_{\rm CO[5-4]}$ relation observed by normal and starburst galaxies with approximately solar metallicity.
 
Although not well characterised to date, evidence that denser molecular gas phases in high-z galaxies can be affected by low metallicity has also been discussed by \mbox{\citet{ref:L.Tacconi2008}} and \citet{ref:R.Genzel2012}, who find increasing numbers of galaxies at z$\sim$2 (e.g. ZC406690, \citealt{ref:C.Mancini2011}) that lack CO[3-2] emission compared to expectations. Our results suggest that this effect is present to even higher-J CO transitions at z=2. We see evidence that the low metallicity in these galaxies may be either directly or indirectly influencing denser molecular gas reservoirs, implying that the linear correlation between CO[5-4] luminosity and SFR (which underpins SFR estimates based on CO emission) may not hold for low-metallicity galaxies at z=2. The deficit of CO emission in these galaxies is particularly relevant when considering high-J CO emission as an instantaneous star-formation tracer (e.g. \citealt{ref:N.Lu2014}, \citealt{ref:E.Daddi2015}, \citealt{ref:D.Liu2015}, \citealt{ref:C.Yang2017}).

\subsection{Metallicity-dependence of \aco\ and G/D for z$\sim$2 galaxies}
\label{sec:Zdependence}

The dependence of \aco\ and G/D on metallicity at z$\sim$0 follows a strong inverse relation, observationally following a power-law Z$^{\gamma}$ or broken power-law shape (e.g. \citealt{ref:DraineLi2007, ref:M.Wolfire2010, ref:A.Leroy2011a, ref:A.RemyRuyer2014}). However, as discussed extensively in the literature (e.g. \citealt{ref:L.Kewley2008}), different metallicity calibrations lead to systematically different gas-phase metallicity derivations (by up to $\sim$0.6~dex). These offsets lead to different normalisations for G/D and \aco\ vs. metallicity trends being reported in the literature (see, e.g., \citealt{ref:K.Sandstrom2013}). For example, at a fixed oxygen abundance value and when assuming a power-law dependence on metallicity with exponent $\gamma$~=~-1, mean G/D ratios (and \aco\ values) will differ by a factor 4.7 when using two of the metallicity scales with among the largest systematic offsets, namely the \citet[][PT05]{ref:PT05} and \citet[][KK04]{ref:KK04} systems respectively. If we assume an even steeper slope with $\gamma$~=~-2, this discrepancy reaches a factor of 22 between the conversion factors derived using the PT05 and KK04 calibrations. It is therefore imperative to correct for the offsets between metallicity scales to ensure meaningful results and comparisons, as these systematics complicate making robust statements about the evolution of scaling relations with redshift.

For this reason, in the following we consider a normalised metallicity scale (see Fig.~\ref{fig:alphaCO_Z}) which minimises the systematic uncertainties introduced by different metallicity scales. We normalise all metallicities by the metallicity Z$_{M_{\star}(MW, z=0)}$ of a z$\sim$0 Milky-Way-like galaxy of stellar mass M$_{\star}$=5.7$\times$10$^{10}$M$_{\odot}$ \mbox{\citep{ref:T.Licquia2015}}, using the relevant local MZR for each calibration \citep{ref:L.Kewley2008}. This is preferable to using a single, constant reference metallicity value, which has conflicting interpretations between different metallicity scales. 
For example, the canonical solar oxygen abundance 12+log$_{10}$(O/H)$_{\odot}$ = 8.69 \citep{ref:M.Asplund2009} is $\sim$0.3~dex higher than the typical MW-like metallicity of Z=8.37 in the PT05 scale. Conversely, on the KK04 scale, the solar abundance value is $\sim$0.3~dex smaller than that of a MW-like galaxy. The advantage of our normalisation approach is thus that we can reconcile scaling relations expressed in different metallicity calibrations by means of our calibration-dependent normalisation factor Z$_{M_{\star}(MW, z=0)}$. If we were to instead convert the range of literature studies to one chosen metallicity system, we would find that straight-line, log-linear scaling relations in one system would map into curved loci in another. Converting to a normalised metallicity scale, as we have done, therefore preserves the shapes of the trend lines, and illustrates the full range of slopes reported in the literature. Additionally, conversions between certain metallicity calibrations are not available in \citet{ref:L.Kewley2008}, which would limit the relations available for comparison in this framework. Common calibrations used among literature relations are PT05, PP04, D02 and KK04, for which Z$_{M_{\star}(MW, z=0)}$ equals 8.37, 8.77, 8.85 and 9.05 respectively as outlined above.

\subsubsection{$\alpha_{\rm CO}$ as a function of metallicity}
\label{sec:aco}
Gas-phase metallicity is observed in the local Universe to have a direct effect on the \aco\ conversion factors of low-metallicity galaxies (e.g. \citealt{ref:A.Leroy2007, ref:R.Amorin2016, ref:G.Accurso2017}, \citealt{ref:L.Hunt2017}). The low gas-phase metallicity strongly increases \aco, as the lower metal content of the ISM leads to lower dust shielding of the CO molecule, which can consequently be photo-dissociated by UV radiation from young stars. This leads to `CO-dark' \Htwo\ gas.

Fig.~\ref{fig:alphaCO_Z} (a) shows $\alpha_{\rm CO}$ as a function of metallicity. For all galaxies in our sample we have placed lower limits on $\alpha_{\rm CO}$ using the gas masses inferred in Section~\ref{sec:gasmass} and the upper limits on the CO[1-0] luminosity listed in Table~\ref{tab:SFRs}:

\begin{equation}
\alpha_{\rm CO} = \frac{M_{mol}}{L^{\prime}_{\rm CO[1-0]}} = \frac{SFR}{SFE}\frac{1}{L^{\prime}_{\rm CO[1-0]}}\\
\end{equation}

We plot in Fig.~\ref{fig:alphaCO_Z} (a) the relations observed by \citet{ref:R.Genzel2012, ref:G.Magdis2012}, \citet{ref:A.Schruba2012} and \citet{ref:K.Sandstrom2013}, and model predictions from \citet{ref:M.Wolfire2010}. The \citet{ref:R.Genzel2012} sample comprises MS galaxies at z$>$1, and \citet{ref:G.Magdis2012} fit local MS spiral galaxies from \citet{ref:A.Leroy2011a} with MS galaxies at 0.5$<$z$<$1.5. \citet{ref:A.Schruba2012} use observations of HERACLES local dwarf galaxies to derive the relation shown, and \citet{ref:K.Sandstrom2013} study local HERACLES star-forming galaxies, the majority of which (25/26) are spiral galaxies.

The 3$\sigma$ lower limits on $\alpha_{\rm CO}$ shown for our sample in Fig.~\ref{fig:alphaCO_Z}~(a) are consistent with the measured trends at low metallicity in the local Universe (e.g \citealt{ref:A.Leroy2011a, ref:A.Schruba2012, ref:K.Sandstrom2013}). Our deeper constraint on \aco\ from stacking (green star in Fig.~\ref{fig:alphaCO_Z} (a)) is also consistent with these low-z trends, as well as the model predictions from \citet{ref:M.Wolfire2010}.

We also show the result of a study at higher redshift, conducted by \citet{ref:R.Genzel2012}. \citet{ref:R.Genzel2012} studied the effect of metallicity on \aco\ including MS galaxies up to z$\sim$2, tracing the molecular gas through the CO[3-2] and CO[2-1] transitions. In order to calculate these conversion factors, \citet{ref:R.Genzel2012} assume a constant gas depletion time for all galaxies, corresponding to a KS slope of n=1.1. The inclusion of our data extends this work, allowing us to study this trend down to a lower median metallicity than the \citet{ref:R.Genzel2012} sample, directly using the ground-state transition CO[1-0]. We find that both our individual and stacked limits are consistent with the z$>$1 \citet{ref:R.Genzel2012} trend. We perform a linear fit between the median \aco\ and metallicity of the z$>$1 galaxies analysed by \citet{ref:R.Genzel2012} (cyan star in Fig.~\ref{fig:alphaCO_Z} (a)) and our stacked limit. We find a lower limit on the dependence with metallicity of $Z^{-0.8}$, which is shallower than the low-redshift studies presented in Fig.~\ref{fig:alphaCO_Z} (a), but consistent with the range of slopes present in the literature (e.g. \citealt{ref:C.Wilson1995, ref:N.Arimoto1996, ref:A.Bolatto2008}). This value does not suggest a steeper slope between \aco\ and metallicity at high redshift than in the local Universe, as \mbox{\citet{ref:R.Genzel2012}} report, but it is based on limited statistics and a low-S/N CO[1-0] flux constraint. It should also be noted that the \citet{ref:R.Genzel2012} sample contains a larger number of galaxies in the range 1$\leq$~z~$\leq$1.5 than at 2~$\leq$~z~$\leq$2.4, and therefore has a lower median redshift than our stacked sample.

To summarise, Fig.~\ref{fig:alphaCO_Z} (a) demonstrates that low-enrichment galaxies at z=2 have higher \aco\ values than their more enriched counterparts, consistent with what is found in the local Universe. We now look at the G/D ratios of these galaxies, to explore the impact of low metallicity at high redshift when deriving gas masses from dust continuum emission.

\subsubsection{Gas-to-dust ratio as a function of metallicity}
\label{sec:GD_Z}
As ISM spectroscopy of galaxies is time-consuming and not always feasible, approaches to using the dust continuum emission of galaxies to estimate gas mass have also been developed (e.g. \citealt{ref:M.Guelin1993, ref:E.Corbelli2012, ref:S.Eales2012, ref:G.Magdis2012, ref:N.Bourne2013}, \citealt{ref:N.Scoville2014, ref:B.Groves2015}). A typical value for the G/D ratio of a `normal', solar-metallicity galaxy in the local Universe is $\sim$100 \citep{ref:DraineLi2007}. We show the G/D ratio as a function of metallicity in Fig.~\ref{fig:alphaCO_Z} (b). We have calculated the dust mass of the galaxies using their continuum emission at 2mm and 870$\micron$ where possible, as discussed in Section~\ref{sec:duststack}. The G/D ratio is calculated as below, where the gas mass derivation is described in Section~\ref{sec:gasmass}:

\begin{equation}
G/D = \frac{M_{mol}+M_{HI}}{M_{d}} \simeq \frac{M_{mol}}{M_{d}} = \frac{SFR}{SFE}\frac{1}{M_{d}}\\
\label{eqn:gd}
\end{equation}

We assume, following standard practice, that the gas phase in our galaxies at z=2 is dominated by molecular gas \citep{ref:E.Daddi2010b, ref:R.Genzel2015, ref:C.dPLagos2015, ref:L.Tacconi2018}. In addition to the individual G/D ratios, we show in Fig.~\ref{fig:alphaCO_Z} (b) the stacked limit on the G/D ratio. It should be noted that Fig.~\ref{fig:alphaCO_Z} (b) shows total gas to dust ratios. In the case of local galaxies, where the molecular gas is not necessarily dominant, the total gas mass therefore also includes atomic hydrogen. The contribution of elements heavier than hydrogen is also accounted for at all redshifts\footnote{We note that the prescriptions of \citet{ref:F.Galliano2011}, adopted by \citet{ref:A.RemyRuyer2014} to derive dust masses, and the one used by \citet{ref:K.Sandstrom2013} and in our work \citep{ref:DraineLi2007} are different. As \citet{ref:K.Sandstrom2013} and \citet{ref:A.RemyRuyer2014} derive G/D ratios for a number of the same KINGFISH galaxies, we renormalise the G/D ratios in \citet{ref:A.RemyRuyer2014} by the mean G/D of the \citet{ref:K.Sandstrom2013} sample, G/D=92.}.

We can use Fig.~\ref{fig:alphaCO_Z} (b) to evaluate whether low-metallicity galaxies at z=2 follow the same scaling relation between G/D and metallicity as local galaxies (e.g. \citealt{ref:K.Sandstrom2013, ref:A.RemyRuyer2014}), and compare our sample with high-z studies such as those by \citet{ref:A.Saintonge2013}. \citet{ref:G.Magdis2012} and \mbox{\citet{ref:K.Sandstrom2013}} use the same galaxy samples described in Section \ref{sec:aco}. \citet{ref:MunozMateos2009} use observations from the local SINGS galaxy sample, and \citet{ref:A.RemyRuyer2014} also study local galaxies, including dwarf galaxies from the Dwarf Galaxy Survey and KINGFISH. \citet{ref:A.Saintonge2013} use observations of $\sim$7 lensed normal star-forming galaxies at z$\sim$2. The best-fit relations derived from each of these works is shown in Fig.~\ref{fig:alphaCO_Z} (b).

With the exception of \citet{ref:MunozMateos2009}, the scaling relations from the literature display an approximately linear decrease of the G/D ratio with metallicity above $\sim$0.5Z$_{norm}$. Below $\sim$0.5Z$_{norm}$, a split occurs, where \citet{ref:G.Magdis2012}, \citet{ref:A.Saintonge2013} and \citet{ref:K.Sandstrom2013} predict a shallower dependance of G/D on Z than \citet{ref:MunozMateos2009} and \citet{ref:A.RemyRuyer2014}. Our stacked sample in particular probes this lower metallicity regime at z=2.

We find that the individual G/D ratios of our low-metallicity galaxies are consistent with all of the local relations shown (noting that all galaxies except V422 are also 2$\sigma$ upper limits on metallicity), as well as the z$>$2 relation from \citet{ref:A.Saintonge2013}. \citet{ref:A.Saintonge2013} report an increased normalisation at z$>$2 of a factor 1.7 compared with the local \citet{ref:A.Leroy2011a} relation. We show the median metallicity and G/D from \citet{ref:A.Saintonge2013} in Fig.~\ref{fig:alphaCO_Z} (b), and find the difference between this value and the \citet{ref:A.Leroy2011a} relation to remain consistent with this factor 1.7 increase also after scaling of all studies to a common, renormalised metallicity scale. We do however note that the local relation of \citet{ref:MunozMateos2009} shown in Fig.~\ref{fig:alphaCO_Z} (b) is also consistent with the \citet{ref:A.Saintonge2013} median data point at z$>$2.

When we consider our stacked G/D limit, we see a clear elevation above the relations presented by \citet{ref:G.Magdis2012}, \citet{ref:A.Saintonge2013} and \citet{ref:K.Sandstrom2013} at this metallicity. We verify through jackknife analysis that this stacked G/D ratio is not strongly driven by the G/D of V580. We strongly favour the case of an increased G/D ratio at significantly sub-solar metallicities, as in seen in our sample. Our stacked limit is in good agreement with both the relations of \citet{ref:MunozMateos2009} and \citet{ref:A.RemyRuyer2014} at this metallicity. Note also that \citet{ref:A.RemyRuyer2014} would derive higher G/D ratios if they were to assume MW-like dust grains, as discussed in \citet{ref:F.Galliano2011}, giving a steeper slope between G/D and metallicity. Fig.~\ref{fig:alphaCO_Z} (b) suggests that our galaxies at z=2 would be consistent with the \citet{ref:A.RemyRuyer2014} relation even for the steeper slope resulting as a consequence of adopting MW-like dust grains. We find a lower limit on the slope between our stacked G/D limit and the median \citet{ref:A.Saintonge2013} value of -2.2, compared with the slopes of -2.45 and -3.10 found by \citet{ref:MunozMateos2009} and \citet{ref:A.RemyRuyer2014} in this metallicity regime, respectively. The calculation of gas masses in our galaxies accounts for the increase in SFE for galaxies above the MS. In order for the lower limit on our G/D vs. Z slope to become consistent with the slope of local relations from \citet{ref:G.Magdis2012} and \citet{ref:K.Sandstrom2013}, the average SFE would need to increase by at least an additional factor of 1.5. The depletion timescale fit given by \citet{ref:L.Tacconi2018}, which suggests a steeper evolution of SFE with offset from the MS than \citet{ref:M.Sargent2014}, as well as a dependence of SFE on stellar mass, is consistent with this order of decrease in G/D limit.

The constraints provided by our data are compatible with a G/D-metallicity relation at z$\sim$2 that does not evolve compared to low redshift. However, they strongly support a steep dependence on metallicity at low enrichment, as shown by \citet{ref:A.RemyRuyer2014}. They are also in agreement with the fudicial model predictions of \citet{ref:G.Popping2017}, which reproduces the double power-law shape observed by \citet{ref:A.RemyRuyer2014}, and predicts only weak evolution of the G/D ratio with redshift ($<$0.3~dex increase in G/D at $<$0.5Z$_{\odot}$ from z=0 to z=9).

It is not trivial to assess how deviations of our galaxies from the star-forming MS (1.6$\sigma$ above, on average) affect the position of these galaxies on scaling relations of G/D and $\alpha_{\rm CO}$ vs. metallicity (sections~\ref{sec:aco} and \ref{sec:GD_Z}). It has been established that galaxies above the MS tend to have larger gas fractions that those galaxies below the MS (e.g. \citealt{ref:G.Magdis2012}, \citealt{ref:A.Saintonge2012}, \citealt{ref:P.Santini2014}, \citealt{ref:M.Sargent2014}, \citealt{ref:N.Scoville2017}, \citealt{ref:L.Tacconi2018}). However, gas-phase metallicity may also correlate with offset from the MS (as given by the Fundamental Metallicity Relation, FMR, \citealt{ref:Lara-Lopez2010b}, \citealt{ref:C.Mannucci2010}), where an increase in SFR at fixed stellar mass implies lower gas-phase metallicity (see e.g. \citealt{ref:R.Sanders2015}, \citealt{ref:M.Onodera2016} and \citealt{ref:G.Cresci2018} for discussion on the existence of an FMR beyond the local Universe). Selection of galaxies with a systematic offset with respect to the MS may therefore not necessarily lead to shifts above and below the scaling relations we present here, but certainly have a component along the metallicity axis of these relations. However, the effect of main-sequence offset on ISM scaling relations cannot be dissected with the number of galaxies studied in this paper.

\subsection{Implications for gas mass derivations at z$\sim$2}
Figs~\ref{fig:alphaCO_Z} (a) and (b) demonstrate that the sub-solar metallicity of our galaxies at z=2 is impacting both the CO-to-H$_{2}$ conversion factor \aco\ and G/D ratio of the galaxies in our sample. Here we explore the implications of this for molecular gas mass calculations in the sub-M$^{*}$ mass regime at the peak epoch of star formation.

\subsubsection{Gas content of individual sub-M$^{*}$ galaxies}

As discussed, the most simplistic approach to measuring gas masses is to assume that \aco\ or G/D do not vary with metallicity. In this case, MW-like conversion factors are often assumed regardless of a galaxy's properties. We find that our stacked \aco\ limit lies a factor $>$2.4 above the MW value \aco=4.4. The deviation from a constant MW value will continue to increase at lower enrichment. Similarly, if we take a constant MW G/D=100, the G/D of our stack implies gas masses $>$4.8 times higher than would be inferred from the MW value.

Another approach one might take to calculate gas masses is to use conversion factors based on a well-studied population of galaxies at a given redshift (e.g., for z$\sim$2, galaxies with stellar masses 10$^{10}$-10$^{11}$M$_{\odot}$). However, galaxies at the knee of the stellar mass function (M$_{\star}\sim$10$^{10.4}$M$_{\odot}$; e.g. \citealt{ref:I.Davidzon2017}) at z$\sim$2 have metallicities approximately $\sim$0.47~dex higher than, e.g., M$_{\star}$$\sim$10$^{9}$M$_{\odot}$ galaxies, depending on the metallicity calibration used \citep{ref:L.Kewley2008, ref:H.Zahid2014a}. Taking a slope of log$_{10}$(\aco) vs. metallicity of -0.8 as suggested by our data would therefore give rise to an underestimation of the gas mass from CO by at least a factor 3, remembering that our data give a lower limit on this slope.
Additionally, there is a steep dependence of G/D on metallicity in this regime ($<$0.5Z$_{M_{\star}(MW, z=0)}$), where G/D $\propto$ Z$^{-2.2}$ or even steeper according to our data. We see in Fig.~\ref{fig:alphaCO_Z} (b) that applying the G/D ratio of an M$_{\star}$$\sim$10$^{10.4}$M$_{\odot}$ galaxy to an M$_{\star}$$\sim$10$^{9}$M$_{\odot}$ galaxy at z=2 would result in an underestimation of the gas mass by a factor $>$17 \citep{ref:H.Zahid2014a}.

These calculations highlight the importance of using physically motivated conversion factors when deriving gas masses for low enrichment or sub-M$^{*}$ galaxies. Assuming MW-like conversion factors or factors appropriate for more massive galaxies would result in an underestimation of the molecular gas mass from both continuum and line emission.

\subsubsection{Which galaxy population dominates the cosmic gas density at z$\sim$2?}

Finally, it is interesting to consider the implications of these findings for measurements of the cold gas history of the universe (e.g. \citealt{ref:C.Lagos2011, ref:G.Popping2014, ref:F.Walter2014, ref:R.Decarli2016, ref:N.Scoville2017, ref:D.Riechers2018}), and in particular for the determination of which characteristic stellar mass scale contributes most to the comoving molecular gas density. Whether - given a set of observed dust masses or molecular line luminosities - we infer this to be, e.g., the M$^{*}$ population or galaxies at lower masses partially depends on how strongly \aco\ and G/D ratios depend on metallicity.
The contribution of galaxies with stellar mass $M_{\star}$ to the comoving molecular gas density is:

\begin{equation}
\begin{split}
\rho_{\rm H_{2}}(M_{\star}) & \propto \Phi(M_{\star}) \langle M_{\rm H_{2}}\rangle_{M_{\star}}\\ & \propto \Phi(M_{\star}) \langle \alpha_{\rm CO} L^{\prime}_{\rm CO[1-0]}\rangle_{M_{\star}}
\end{split}
\end{equation}

(or $\rho_{\rm H_2}(M_{\star}) \propto\Phi(M_{\star}) \langle G/D \times M_d\rangle_{M_{\star}}$). Here $\langle M_{\rm H_2}\rangle_{M_{\star}}$ is the characteristic H$_{2}$ mass of galaxies with stellar mass M$_{\star}$, and $\Phi(M_{\star}$) is the comoving number density of galaxies with a given stellar mass, e.g. from the stellar mass function of actively star-forming galaxies in \citet{ref:I.Davidzon2017}. For log$_{10}$(M$_{\star}$)~=~9 (10.4) at z$\sim$2, log$_{10}$($\Phi(M_{\star})$)~=~-2.26 (-2.86), corresponding to a 4-fold higher number density of M$_{\star}$$\sim$10$^{9}$M$_{\odot}$ systems, but a $\sim$6-fold higher contribution of 10$^{10.4}$M$_{\odot}$ galaxies to the comoving stellar mass density $\rho(M_{\star})$. By how much will the larger \aco\ values for sub-M$^{*}$ galaxies modify the relative contribution of these two stellar mass bins in terms of the z$\sim$2 cosmic gas mass density (compared to their contribution to the stellar mass density)?

Using our derived lower limits on the slopes of the log$_{10}$(\aco) and log$_{10}$(G/D) vs. Z relations, we can derive a lower bound on the relative contribution to the cosmic gas mass density of galaxies at M$_{\star}$=10$^{9}$M$_{\odot}$, and at the knee of the stellar mass function at M$^{*}$=10$^{10.4}$M$_{\odot}$. We find an increased ratio $\frac{\rho_{H_{2}}(M_{\star}=10^{9}M_{\odot})}{\rho_{H_{2}}(M_{\star}=10^{10.4}M_{\odot})}$$>$11$\times$$\frac{L^{\prime}_{\rm CO[1-0]}(M_{\star}=10^{9}M_{\odot})}{L^{\prime}_{\rm CO[1-0]}(M_{\star}=10^{10.4}M_{\odot})}$ of the comoving gas densities when using a metallicity dependant \aco, compared with the considerably smaller increase of 4 times the L$^{\prime}_{\rm CO[1-0]}$ ratio when a constant MW-like \aco\ is assumed. When a metallicity dependent G/D is used, we find an increase in the comoving gas density of $\frac{\rho_{H_{2}}(M_{\star}=10^{9}M_{\odot})}{\rho_{H_{2}}(M_{\star}=10^{10.4}M_{\odot})}$$>$68$\times$$\frac{M_{d}(M_{\star}=10^{9}M_{\odot})}{M_{d}(M_{\star}=10^{10.4}M_{\odot})}$, compared to just 4 times the dust mass ratios when a constant G/D is assumed. For the contribution to the z$\sim$2 cosmic gas mass density of sub-M$^{*}$ galaxies, $\rho_{H_{2}}(M_{\star}=10^{9}M_{\odot}$), to become at least as large as that of galaxies at the knee of the stellar mass function, the ratio of the typical L$^{\prime}_{\rm CO[1-0]}$ (M$_{d}$) values in the low-mass bin over that in the high-mass bin would thus have to be $\geq$0.09 ($\geq$0.015). If we follow the same formalism as in Section~\ref{sec:lco} to relate SFR, M$_{\star}$, M$_{\rm H_{2}}$ and Z for average galaxies in these mass bins (namely using the star-formation MS, the MZR and the integrated KS law to find SFR, metallicity and molecular gas mass, for a given M$_{\star}$), and take a power law index $\gamma$=-0.8 of \aco\ vs. metallicity from our data to derive \aco, we can estimate average L$^{\prime}_{\rm CO[1-0]}$ values for these stellar mass bins at z=2 (using L$^{\prime}_{\rm CO[1-0]}$=M$_{\rm H_{2}}$/\aco). We predict the average L$^{\prime}_{\rm CO[1-0]}$ ratio to be 0.045, indicating that despite the increased \aco\ in sub-M$^{*}$ galaxies, the population at M$_{\star}$=10$^{9}$M$_{\odot}$ has not overtaken that at M$^{*}$=10$^{10.4}$M$_{\odot}$ in terms of their contributions to the comoving gas mass density at z$\sim$2. This also implies a much lower CO detection rate for low mass galaxies at z=2, compared with more massive, enriched galaxies.

Rising CO-to-H$_{2}$ conversion factors and G/D ratios at low mass/metallicity do, however, significantly shift the balance between different galaxy mass scales compared to the stellar mass density distribution. For example, in our example involving a metallicity-dependent \aco, a more than 6-fold higher contribution of M$^{*}$ galaxies to $\rho$(M$_{\star}$) was reduced to an only 2$\times$ higher contribution to $\rho$(M$_{H_2}$), compared to the M$_{\star}$=10$^{9}$M$_{\odot}$ population.

\section{Conclusions}
\label{sec:conc}
This paper explores the relationship between molecular gas, dust, star-formation and metallicity in five sub-solar enrichment galaxies (average oxygen abundance $\sim$0.35~dex below solar) at z$\sim$2. Studying the ISM of these sub-M$^{*}$ galaxies gives us a greater understanding of how low-metallicity ISM scaling relations at z=2 compare with those calibrated in the local Universe, and on more massive and luminous z$\sim$2 galaxies. Limits on the oxygen abundance of these galaxies were first constrained by \citet{ref:F.Valentino2015}, making use of the galaxies' bright \Halpha\ detections. Here we follow the same procedure to derive a sample average metallicity of Z=8.34$\pm$0.13 (PP04 N2 calibration).

Using deep ALMA and VLA data, we search for the CO[1-0] and CO[4-3] lines and 870$\micron$ and 2mm continuum emission from our targets, but find that they remain individually undetected. These non-detections highlight the low ISM enrichment of our galaxies, particularly when comparing with the expected detection percentage (60-80\%) in our data, for solar enrichment galaxies with these \Halpha\ SFRs. We perform stacked analysis on the CO spectra and dust emission in order to place more stringent limits on the average ISM content of the five galaxies, and once again find these quantities undetected in the stacked data.

We place our galaxies on the L$^{\prime}_{\rm CO[1-0]}$-SFR plane, and see that our sample average has an offset of $>$0.5~dex below the expected L$^{\prime}_{\rm CO[1-0]}$, occupying the same parameter space as high-SFE starburst galaxies. We verify that this offset is not driven by a recent increase in star-formation, nor a high SFE in our galaxies. We therefore interpret this as evidence that the low metal enrichment of the ISM in our galaxies at z=2 is leading to dissociation of CO in the diffuse ISM phase, as seen in the local Universe.

This CO deficit also manifests itself for the denser ISM phases which are traced by CO[4-3]. We convert our stacked CO[4-3] constraint to CO[5-4] luminosity using a BzK excitation correction, in order to examine the position of our galaxies on the L$^{\prime}_{\rm CO[5-4]}$-SFR plane. We find an average luminosity of $>$0.5-0.7~dex less than the expected CO[5-4] luminosity. The position of galaxies on this relation is invariant to their star-formation efficiency, confirming that this CO deficit is most likely driven by the low gas-phase metallicity. This has important implications, as it could lead to systematic underestimation of the instantaneous SFRs of low-metallicity galaxies as traced through high-J CO transitions. 

In order to investigate how the ISM properties of our galaxies scale with metallicity at z=2, we calculate molecular gas masses using the \Halpha\ SFRs, assuming a MS SFE. We then derive limits on the \aco\ conversion factors and G/D ratios of the galaxies. We find a power law slope of \aco\ (G/D) vs. metallicity at z$\sim$2 of $\gamma<-0.8$ ($\gamma<-2.2$), consistent with dependencies observed at z=0. We stress the importance of comparing studies using different metallicity calibrations on a normalised metallicity scale, and place a compilation of literature scaling relations alongside our data. When systematic differences in metallicity are taken into account in this way, we find consistency between both the \aco\ and G/D vs. metallicity relations at z=2 compared with local relations. However, our stacked G/D limit strongly favours a steep G/D vs. Z slope at low-metallicity ($\sim$0.5Z$_{M_{\star}(MW, z=0)}$), consistent with the double-power law relation given by \citet{ref:A.RemyRuyer2014}.

Finally, we discuss the implications of these results on the calculation of molecular gas masses in low-metallicity, z=2 galaxies. We find that the assumption of G/D and \aco\ values based on the more routinely observed M$^{*}$ population at this redshift (or uniform application of the `canonical' MW-like conversion factors) would result in significant underestimation of the gas mass in these low-metallicity galaxies. We predict the average contribution of sub-M$^{*}$ galaxies with M$_{\star}$=10$^{9}$M$_{\odot}$ to the comoving molecular gas density at z=2, and find that their increased \aco\ values are not sufficient for this population to become dominant over M$^{*}$ galaxies in this respect. However, we see that increased \aco\ and G/D conversion factors at low enrichment significantly shift the balance between different galaxy mass scales compared to their relative contribution to stellar mass density distributions.

Our analysis has enabled us to place observational constraints on the ISM properties of low-enrichment galaxies at the peak epoch of star-formation. In combination with future, similar studies, this kind of work can inform the design of ISM surveys of sub-M$^{*}$ galaxies. Going forward, collating larger, statistical samples of galaxies spanning a large metallicity range will allow us to determine, e.g. the influence of starburstiness or MS-offset on the scatter of the ISM scaling relations studied here. To our knowledge, this has not yet been quantified at low or high redshift. In this way we will continue to explore the effect of low gas-phase metallicity on the interstellar medium, and characterise the evolution of the ISM outside of the local Universe.

\section*{Acknowledgements}

We thank the anonymous referee, whose comments improved the clarity of this paper. RTC acknowledges support from the STFC (grant ST/N504452/1) and the University of Sussex. MTS was supported by a Royal Society Leverhulme Trust Senior Research Fellowship (LT150041).

This paper makes use of the following ALMA data: ADS/JAO.ALMA\#2012.1.00885.S and ADS/JAO.ALMA\#2015.1.01355.S. ALMA is a partnership of ESO (representing its member states), NSF (USA) and NINS (Japan), together with NRC (Canada), MOST and ASIAA (Taiwan), and KASI (Republic of Korea), in cooperation with the Republic of Chile. The Joint ALMA Observatory is operated by ESO, AUI/NRAO and NAOJ.

This paper also makes use of JVLA program 12A-188. The National Radio Astronomy Observatory is a facility of the National Science Foundation operated under cooperative agreement by Associated Universities, Inc.





\bibliographystyle{mnras}
\bibliography{EnvironmentNew} 





\bsp	
\label{lastpage}
\end{document}